\documentstyle[12pt,epsfig]{article}
\def\bge{\begin{equation}}
\def\ene{\end{equation}}
\def\bg{\begin{eqnarray}}
\def\en{\end{eqnarray}}
\def\nn{\nonumber}

\def\Dbar{\overline{D}}
\def\D0bar{\overline{D^0}}

\begin{document}
\begin{titlepage}
\title{
$\Lambda^+_c$, $\Sigma_c$, $\Xi_c$ and $\Lambda_b$ hypernuclei\\
in the quark-meson coupling model
}  
\author{\\
K  Tsushima$^{1,2}$~\thanks{tsushima@ift.unesp.br}~ and 
F C  Khanna$^{3,4}$~\thanks{khanna@Phys.UAlberta.CA}
\\ \\
{$^1$\small Instituto de F\'{\i}sica Te\'orica - UNESP, 
Rua Pamplona 145, 01405-900, Sao Paulo - SP, Brazil}\\
{$^2$\small Mackenzie University - FCBEE, Rua da Consolacao 930,
01302-907, Sao Paulo - SP, Brazil 
} \\
{$^3$\small Department of Physics, University of Alberta, Edmonton, 
Canada, T6G 2J1} \\
{$^4$\small TRIUMF, 4004 Wesbrook Mall, Vancouver, British Columbia, 
Canada, V6T 2A3}
}
\date{}
\maketitle
\vspace{-9.5cm}
\hfill Alberta THy 07-03
\vspace{9.5cm}
\begin{abstract}
Charmed (and bottom) hypernuclei are 
studied in the quark-meson coupling (QMC) model.
This completes systematic studies of  
charmed ($\Lambda_c^+, \Sigma_c, \Xi_c$), and $\Lambda_b$ 
hypernuclei in the QMC model.
Effects of the Pauli blocking due to 
the underlying quark structure of baryons, 
and the $\Sigma_c N - \Lambda_c N$ channel 
coupling are phenomenologically taken into account   
at the hadronic level in the same way as 
those included for strange hypernuclei. 
Our results suggest that the $\Sigma_c^{++}$ and 
$\Xi_c^+$ hypernuclei are very unlikely to be formed, 
while the $\Lambda_c^+$, $\Xi_c^0$ and $\Lambda_b$ 
hypernuclei are quite likely to be formed.
For the $\Sigma_c^+$ hypernuclei, 
the formation probability is non-zero though small.
A detailed analysis is also made about the phenomenologically 
introduced Pauli blocking and channel coupling effects for 
the $\Sigma_c^0$ hypernuclei.
\end{abstract}
\end{titlepage}
%
%
\section{Introduction}

Based on the quark-meson coupling (QMC) model~\cite{Guichon}, 
we have recently initiated a systematic study of 
charmed and bottom hadrons in nuclear matter~\cite{TK}.
The results imply that the formations of 
$B^-$-meson nuclear (atomic) bound states, 
and $\Lambda_c^+$ and $\Lambda_b$ hypernuclei  
are quite likely. The formations of 
$\Lambda_c^+$ and/or $\Lambda_b$ hypernuclei 
were first predicted in mid-1970s by Tyapkin~\cite{Tyapkin}, 
and Dover and Kahana~\cite{Dover}. 
Later on theoretical studies~\cite{Gibson,Bando} 
as well as possibility of experimental observation~\cite{bcexp} were made.
Further, we have studied~\cite{TKhypbc} the 
$\Lambda^+_c$ and $\Lambda_b$  
hypernuclei by solving equations of motion for 
a system of hypernuclei, embedding a $\Lambda^+_c$ or a $\Lambda_b$
into the closed shell nuclear core, within the Hartree, mean-field, 
approximation using the QMC model. The results again support  
the possible formation of $\Lambda_c^+$ and $\Lambda_b$ hypernuclei.  

In this article we present the results  
for the $\Sigma_c$ and $\Xi_c$ hypernuclei, as well as  
some of the $\Lambda^+_c$ and $\Lambda_b$ 
hypernuclei studied recently~\cite{TKhypbc}.
This is an extension of the studies made for  
the strange hypernuclei~\cite{Tsushima_hyp}, 
and a completion of our systematic study 
of charmed and bottom hypernuclei~\cite{TK,TKhypbc} 
in the QMC model. 
The extension is straightforward,   
since the heavy baryons (as well as nucleon) are classified by 
the naive constituent quark model, and QMC can treat them 
on the same quark-based footing.

One of the other important issues in this study is a role 
of the $u$ and $d$ light quarks in nuclear medium,  
in the charmed or bottom baryons containing $u$ and/or $d$ quarks. 
This concerns the partial restoration of chiral symmetry in nuclear 
medium for heavy baryons. 

Concerning the experimental possibilities, the approved construction
of the Japan Hadron Facility (JHF)
with a beam energy of 50 GeV, will produce charmed hadrons
profusely and bottom hadrons in lesser
numbers, but still with an intensity which is comparable
to the present hyperon production rates.
Although the study of possible existence of
charmed and bottom hypernuclei may be in its infancy at present,
it is clear that conditions for the experiments to search for such
heavy baryon hypernuclei are now becoming realistic, and would be
realized at JHF~\cite{JHF}.
Thus, it is still meaningful to consider such heavy baryon 
hypernuclei at this stage for the future experimental studies. 

The QMC model (QMC-I), which is used in this study, has been successfully   
applied to many problems of nuclear physics, such as  
finite nuclei~\cite{Guichonf,Saitof}, 
strange hypernuclei~\cite{Tsushima_hyp},
nuclear matter~\cite{matter}, and other problems 
including hadronic properties in nuclear 
medium~\cite{QMCapp,Tsushimaeta,TsushimaD,TsushimaJ,Alexjpsi,Alexd,TsushimaK}.
(See also Refs.~\cite{blu,jin} for earlier, different versions of 
QMC, and Ref.~\cite{QMCinitial} for related work.)
For example, the model was applied to the study of
meson nuclear bound states~\cite{Tsushimaeta,TsushimaD}, 
$J/\Psi$ dissociation in nuclear matter~\cite{Alexjpsi}, 
$D$ and $\Dbar$ production in antiproton-nucleus
collisions~\cite{Alexd}, and kaon properties in nuclear matter and 
kaon production in heavy 
ion collisions~\cite{TsushimaK}.
Furthermore, although only some limited studies for heavy mesons
(not for heavy baryons) with charm
in nuclear matter were made by the QCD sum rules 
(for $J/\Psi$~\cite{Hayashigaki1,Klingl} and
$D (\Dbar)$~\cite{Hayashigaki2}),
a study for heavy baryons containing charm
or bottom quarks was made only recently~\cite{TK} using the QMC model.
Furthermore, recent measurements of polarization transfer in the 
$^4$He($\vec{e},e'\vec{p}$)$^3$H and $^{16}$O($\vec{e},e'\vec{p}$)$^{15}$N 
reactions performed at MAMI and Jefferson Lab~\cite{JlabQMC}, 
support the medium modification of the
proton electromagnetic form factors predicted by the QMC 
model~\cite{TsushimaJ}. 

There is also another version of the QMC model (QMC-II), 
where masses of the meson fields
are also subject to the medium modification in a self-consistent
manner~\cite{qmc2}. However, for a proper
parameter set (set B) the typical results obtained
in QMC-II are very similar to those of QMC-I.
The difference is $\sim 16$ \% for the largest case, but
typically $\sim 10$ \% or less. (For the effective
masses of the hyperons, the differences between the two versions of QMC 
turn out to be less than $\sim 8$ \%. See also 
Fig.~2 in Ref.~\cite{Tsushimaeta} for the differences in the scalar 
potentials for the $\eta$ and $\omega$ mesons in $^{208}$Pb nucleus.)
Since we use QMC-I, the differences in these numbers may be regarded as   
uncertainties in the model.

Certainly, the model has shortcomings that have to be improved eventually.
Difficulties to handle the model will increase rapidly
in handling the Hartree-Fock approximation even for nuclear
matter~\cite{QMCfock}, and further to include the Pauli blocking
effect at the quark level. 
In addition, the $\Sigma N - \Lambda N$ channel
coupling effect has not been implemented yet in 
a consistent manner with
the underlying quark degrees of freedom~\cite{Tsushima_hyp}.
It should be mentioned that in the case of larger mass number
$\Sigma$ hypernuclei,  
no narrow states have been observed experimentally, 
although $^4_{\Sigma}$He was confirmed~\cite{sigmahyp}. 
It is unlikely 
to find such (narrow) bound states in the present situation, 
where the experimental analysis~\cite{sigmahyp} supports the suggestion 
made by Harada~\cite{Harada}, that a large isospin dependent 
$\Sigma$-nucleus potential term exists and the 1/A (A:baryon number) 
dependence of that term reduces the likelihood of observing 
bound states for $A > 5$.
This unlikeliness may also arise from the width of $\Sigma$ state 
due to strong $\Sigma - \Lambda$ conversion.
Furthermore, an application to double hypernuclei has not     
been attempted, although recently the existence
was confirmed for small baryon number nuclei~\cite{doublehyp}.
As an attempt to improve the model, 
the effect of short-range quark-quark
correlations associated with the nucleon overlap 
was studied in Ref.~\cite{Saito_qq}.
With the addition of the correlations, the saturation curve for
symmetric nuclear matter was greatly improved at high density. 
Despite the shortcomings of the model described above, 
there are some positive
aspects of this model, its simplicity and successful
applicability, that we feel confident that the QMC model will
provide us with a valuable glimpse into the
properties of charmed and bottom hypernuclei.

In the QMC model, the interactions between nucleons  
are mediated by the exchange of scalar ($\sigma$) and vector 
($\omega$ and $\rho$) fields self-consistently coupled directly  
to the quarks within those nucleons, but not to the nucleons 
as in Quantum Hadrodynamics (QHD)~\cite{QHD}. 
Thus, it can be systematically applied to study the properties 
of any hadrons   
in nuclear medium~\cite{TK,QMCapp} (if they contain 
$u$ and/or $d$ light quarks).
We make use of this advantage of the QMC model, 
and study the charmed and bottom hypernuclei.
We hope that the present study inspires 
some interest to measure the properties of heavy baryons 
in nuclear medium, at facilities like JHF, particularly  
at RHIC (high energy relativistic heavy ion facilities), and in very high 
energy experiments at CERN and Fermilab.

The organization of this article is as follows. 
In Section~2, the relativistic formulation of the 
charmed and bottom hypernuclear system in the QMC model 
will be explained briefly. A mean-field Lagrangian density and equations of 
motion will be given in Section~2.1, while charmed and bottom hadrons in 
nuclear matter will be discussed in Section~2.2.
In Section~3, the effects necessary for a 
realistic calculation for the charmed (bottom) hypernuclei will be 
described. Spin-orbit potential, and the effect of the Pauli blocking 
at the quark level will be explained in Sections~3.1 and 3.2, respectively. 
In addition, $\Sigma_{c,b} N - \Lambda_{c,b} N$ channel coupling effect  
will be explained in Section~3.2. The results for 
the charmed and bottom hypernuclei 
will be presented in Section~4, and finally 
Section~5 will be devoted to summary and discussion.

%
%
\section{Charmed and bottom hypernuclei in the QMC model}

In this Section, we briefly explain the mean-field equations of motion for 
a charmed or bottom hypernuclear system, and review some properties 
of the charmed and bottom baryons in nuclear matter.

\subsection{Mean-field equations of motion}

We consider static, approximately spherically symmetric
charmed and bottom hypernuclei, i.e., closed shell nuclear core 
plus one charmed or one bottom baryon (heavy baryon) configuration, 
ignoring non-spherical
effects due to the embedded heavy baryon.
The existence of the heavy baryon 
inside or outside of the nuclear core (in particular, $\Lambda_b$ 
which interacts relatively strong with core nucleons) 
breaks spherical symmetry, and one should include this effect 
in a truly rigorous treatment. We have neglected this effect, 
since it is not expected to be so important 
for spectroscopic calculations~\cite{fur,coh} 
for the middle and/or large baryon number charmed hypernuclei
(but may not be true for $\Lambda_b$ hypernuclei), 
and beyond the scope of present study.
However, we include the response of the nuclear core
arising from the self-consistent 
calculation, which is a purely relativistic effect~\cite{coh,coh2}.
Thus, we always specify the
state of the heavy baryon in which the calculation is performed, 
because the core nucleus response is different  
due to the self-consistent calculation.
(This self-consistent procedure will be applied also when 
the effective Pauli blocking 
and/or $\Sigma_{c,b} - \Lambda_{c,b}$ are introduced later.)

We adopt the Hartree, mean-field, approximation.
In this approximation the $\rho NN$ tensor coupling contributes to 
a spin-orbit force for a nucleon bound 
in a static spherical nucleus, although
in Hartree-Fock it also contributes a central 
force~\cite{Guichonf,Saitof}. 
(Furthermore, it gives no contribution for nuclear
matter since the meson fields are independent of position
and time.) Thus, we ignore the $\rho NN$ tensor coupling in this
study as in Ref.~\cite{Tsushima_hyp}. 

A relativistic Lagrangian density for 
a charmed or a bottom hypernucleus  
in the QMC model may be given by~\cite{Tsushima_hyp}:  
\begin{eqnarray}
{\cal L}^{CHY}_{QMC} &=& {\cal L}_{QMC} + {\cal L}^C_{QMC},
\nn \\ 
{\cal L}_{QMC} &=&  \overline{\psi}_N(\vec{r})
\left[ i \gamma \cdot \partial
- M_N^\star(\sigma) - (\, g_\omega \omega(\vec{r})
+ g_\rho \frac{\tau^N_3}{2} b(\vec{r})
+ \frac{e}{2} (1+\tau^N_3) A(\vec{r}) \,) \gamma_0
\right] \psi_N(\vec{r}) \quad \nn \\
  & & - \frac{1}{2}[ (\nabla \sigma(\vec{r}))^2 +
m_{\sigma}^2 \sigma(\vec{r})^2 ]
+ \frac{1}{2}[ (\nabla \omega(\vec{r}))^2 + m_{\omega}^2
\omega(\vec{r})^2 ] \nn \\
 & & + \frac{1}{2}[ (\nabla b(\vec{r}))^2 + m_{\rho}^2 b(\vec{r})^2 ]
+ \frac{1}{2} (\nabla A(\vec{r}))^2, \nn \\
{\cal L}^C_{QMC} &=& 
\sum_{C=\Lambda^+_c,\Sigma_c^{0,+,++},\atop \Xi_c^{0,+},\Lambda_b}
\overline{\psi}_C(\vec{r})
\left[ i \gamma \cdot \partial
- M_C^\star(\sigma)
- (\, g^C_\omega \omega(\vec{r})
+ g^C_\rho I^C_3 b(\vec{r})
+ e Q_C A(\vec{r}) \,) \gamma_0
\right] \psi_C(\vec{r}), \qquad \label{Lagrangian}
\end{eqnarray}
where $\psi_N(\vec{r})$ ($\psi_C(\vec{r})$)
and $b(\vec{r})$ are respectively the
nucleon (charmed or bottom baryon) and the $\rho$
meson (the time component in the third direction of
isospin) fields, while $m_\sigma$, $m_\omega$ and $m_{\rho}$ are
the masses of the $\sigma$, $\omega$ and $\rho$ meson fields.
$g_\omega$ and $g_{\rho}$ are the $\omega$-$N$ and $\rho$-$N$
coupling constants which are respectively related to 
($u,d$)-quark-$\omega$, $g_\omega^q$, and
($u,d$)-quark-$\rho$, $g_\rho^q$, coupling constants as
$g_\omega = 3 g_\omega^q$ and
$g_\rho = g_\rho^q$~\cite{Guichonf,Saitof}.
Note that these meson fields, $\sigma, \omega$ and $\rho$, represent
the quantum numbers and Lorentz structure which mediate the interactions
among the nucleons in nuclear medium as in original
QHD~\cite{QHD}, corresponding, 
$\sigma \leftrightarrow \phi_0$, $\omega \leftrightarrow V_0$ and 
$b \leftrightarrow b_0$, and they may not be directly connected with the
physical particles, nor quark model states.
Their masses in nuclear medium do not vary in the present treatment of 
QMC-I. Hereafter we will use notations for the quark flavors,
$q \equiv u,d$ and $Q \equiv s,c,b$.

In an approximation where the $\sigma$, $\omega$ and $\rho$ fields couple
only to the $u$ and $d$ quarks,
the coupling constants for the heavy baryon 
are obtained as $g^C_\omega = (n_q/3) g_\omega$, and
$g^C_\rho \equiv g_\rho = g_\rho^q$, with $n_q$ being the total number of
valence $u$ and $d$ light quarks in the heavy baryon $C$. $I^C_3$ and $Q_C$
are the third component of the heavy baryon isospin operator and its electric
charge in units of the proton charge, $e$, respectively.
The field dependent $\sigma$-$N$ and $\sigma$-$C$
coupling strengths predicted by the QMC model,
$g_\sigma(\sigma) \equiv g^N_\sigma(\sigma)$ and  $g^C_\sigma(\sigma)$,
related to the Lagrangian density of 
Eq.~(\ref{Lagrangian}) are defined by 
\bg
M_N^\star(\sigma) &\equiv& M_N - g_\sigma(\sigma)
\sigma(\vec{r}) ,  \\
M_C^\star(\sigma) &\equiv& M_C - g^C_\sigma(\sigma)
\sigma(\vec{r}) , \label{effective_mass}
\en
where $M_N$ ($M_C$) is the free nucleon (heavy baryon) mass.
Note that the dependence of these coupling strengths on the applied
scalar field must be calculated self-consistently within the quark
model~\cite{Tsushima_hyp,Guichonf,Saitof}.
Hence, unlike QHD~\cite{QHD}, even though
$g^C_\sigma(\sigma) / g_\sigma(\sigma)$ may be
2/3 or 1/3 depending on the number of light quarks in the heavy baryon
in free space, $\sigma = 0$, (but only when their bag radii in free space 
are exactly the same), 
this will not necessarily  be the case in
nuclear matter.
Explicit expressions for $g^C_\sigma(\sigma)$
and $g_\sigma(\sigma)$ will be given later in Eq.~(\ref{Csigma}). From
the Lagrangian density Eq.~(\ref{Lagrangian}), a set of
equations of motion for the heavy baryon hypernuclear system is obtained: 
\begin{eqnarray}
& &[i\gamma \cdot \partial -M^\star_N(\sigma)-
(\, g_\omega \omega(\vec{r}) + g_\rho \frac{\tau^N_3}{2} b(\vec{r})
 + \frac{e}{2} (1+\tau^N_3) A(\vec{r}) \,)
\gamma_0 ] \psi_N(\vec{r}) = 0, \label{eqdiracn}\\
& &[i\gamma \cdot \partial - M^\star_C(\sigma)-
(\, g^C_\omega \omega(\vec{r}) + g_\rho I^C_3 b(\vec{r})
+ e Q_C A(\vec{r}) \,)
\gamma_0 ] \psi_C(\vec{r}) = 0, \label{eqdiracy}\\
& &(-\nabla^2_r+m^2_\sigma)\sigma(\vec{r}) =
- [\frac{\partial M_N^\star(\sigma)}{\partial \sigma}]\rho_s(\vec{r})
- [\frac{\partial M_C^\star(\sigma)}{\partial \sigma}]\rho^C_s(\vec{r}),
\nn \\
& & \hspace{7.5em} \equiv g_\sigma C_N(\sigma) \rho_s(\vec{r})
    + g^C_\sigma C_C(\sigma) \rho^C_s(\vec{r}) , \label{eqsigma}\\
& &(-\nabla^2_r+m^2_\omega) \omega(\vec{r}) =
g_\omega \rho_B(\vec{r}) + g^C_\omega
\rho^C_B(\vec{r}) ,\label{eqomega}\\
& &(-\nabla^2_r+m^2_\rho) b(\vec{r}) =
\frac{g_\rho}{2}\rho_3(\vec{r}) + g^C_\rho I^C_3 \rho^C_B(\vec{r}),
 \label{eqrho}\\
& &(-\nabla^2_r) A(\vec{r}) =
e \rho_p(\vec{r})
+ e Q_C \rho^C_B(\vec{r}) ,\label{eqcoulomb}
\end{eqnarray}
where, $\rho_s(\vec{r})$ ($\rho^C_s(\vec{r})$), $\rho_B(\vec{r})$
($\rho^C_B(\vec{r})$), $\rho_3(\vec{r})$ and
$\rho_p(\vec{r})$ are the scalar, baryon, third component of isovector,
and proton densities at the position $\vec{r}$ in
the heavy baryon hypernucleus~\cite{Tsushima_hyp,Guichonf,Saitof}.
On the right hand side of Eq.~(\ref{eqsigma}),
$- [{\partial M_N^\star(\sigma)}/{\partial \sigma}] \equiv
g_\sigma C_N(\sigma)$ and
$- [{\partial M_C^\star(\sigma)}/{\partial \sigma}] \equiv 
g^C_\sigma C_C(\sigma)$, where $g_\sigma \equiv g_\sigma (\sigma=0)$ and
$g^C_\sigma \equiv g^C_\sigma (\sigma=0)$,
are a new, and characteristic feature of QMC
beyond QHD~\cite{QHD,mar,ruf,Jennings}.
The effective mass for the heavy baryon $C$ is defined by 
\begin{equation}
\frac{\partial M_C^\star(\sigma)}{\partial \sigma}
= - n_q g_{\sigma}^q \int_{bag} d^3 x 
\ {\overline \psi}_q(\vec{x}) \psi_q(\vec{x})
\equiv - n_q g_{\sigma}^q S_C(\sigma) = - \frac{\partial}{\partial \sigma}
\left[ g^C_\sigma(\sigma) \sigma \right],
\label{gsigma}
\end{equation}
with the MIT bag model quantities, and the mass stability condition 
to be satisfied in obtaining the 
in-medium bag radius~\cite{Guichon,Tsushima_hyp,Guichonf,Saitof}:  
\begin{eqnarray}
& &M_C^\star(\sigma) =
\sum_{j=q,Q}\frac{n_j\Omega^*_j -  z_C}{R_C^*}
+ \frac{4}{3}\pi ({R_C^*})^3 B ,
\label{bagmass} \\
& &S_C(\sigma) = \frac{\Omega_q^*/2
+ m_q^*R_C^*(\Omega_q^*-1)}
{\Omega_q^*(\Omega_q^*-1)
+ m_q^*R_C^*/2},
\label{Ssigma} \\
& &
\Omega_q^* = \sqrt{x_q^2 + (R_C^* m_q^*)^2},
\quad \Omega_Q^* = \sqrt{x_Q^2 + (R_C^* m_Q)^2},\quad
m_q^* = m_q - g_{\sigma}^q \sigma (\vec{r}), 
\label{OmegaqQ} \\
& &C_C(\sigma) = S_C(\sigma)/S_C(0), \quad
g^C_{\sigma} \equiv n_q g_{\sigma}^q S_C(0)
\equiv \frac{n_q}{3}g_\sigma \Gamma_{C/N} 
= \frac{n_q}{3} g_\sigma S_C(0)/S_N(0), 
\label{Csigma}\\
& & \left. \frac{dM^\star_C}{dR_C}
\right|_{{R_C} = R^*_C} = 0.
\label{bagradii}
\end{eqnarray}
Quantities for the nucleon are similarly obtained by replacing the
indices, $C \to N$, in Eqs.~(\ref{gsigma})~-~(\ref{bagradii}).
Approximating the constant, mean meson fields 
within the bag neglecting the Coulomb force, 
the wave functions, $\psi_f (x) (f=u,d,s,c,b)$, 
satisfy the Dirac equations for 
the quarks in the baryon bag 
centered at a position $\vec{r}$ of the nucleus, 
($|\vec{x} - \vec{r}|\le$ 
bag radius~\cite{Tsushimaeta,TsushimaD,TsushimaK}): 
\begin{eqnarray}
\left[ i \gamma \cdot \partial_x -
(m_q - V^q_\sigma(\vec{r}))
- \gamma^0
\left( V^q_\omega(\vec{r}) \pm
\frac{1}{2} V^q_\rho(\vec{r})
\right) \right]
\left( \begin{array}{c} \psi_u(x)  \\
\psi_d(x) \\ \end{array} \right) &=& 0,
\label{diracu}\\
%
%
\left[ i \gamma \cdot \partial_x - m_{Q} \right]
\psi_{Q} (x)\,\, &=& 0, 
\label{diracQ}
\end{eqnarray}
where, $q = u$ or $d$, and $Q = s,c$ or $b$. 
The constant, mean-field potentials within the bag 
are defined by $V^q_\sigma(\vec{r}) \equiv g^q_\sigma \sigma(\vec{r})$,
$V^q_\omega(\vec{r}) \equiv g^q_\omega \omega(\vec{r})$ and
$V^q_\rho(\vec{r}) \equiv g^q_\rho b(\vec{r})$,
with $g^q_\sigma$, $g^q_\omega$ and
$g^q_\rho$ the corresponding quark-meson coupling constants.
The eigenenergies, $\epsilon_{u,d,Q}$  
in units of $1/R_C^*$ are given by 
\bge
\left( \begin{array}{c}
\epsilon_u \\
\epsilon_d
\end{array} \right)
= \Omega_q^* + R_C^* \left(
V^q_\omega(\vec{r})
\pm \frac{1}{2} V^q_\rho(\vec{r}) \right),\hspace{1em} 
%
%
\epsilon_Q = \Omega_Q^*,
\label{energy}
\ene
where, $\Omega_q^*$ and $\Omega_Q^*$ are given in Eq.~(\ref{OmegaqQ}).

In the expressions of the MIT bag quantities 
Eqs.~(\ref{bagmass}) and~(\ref{OmegaqQ}), $z_C$, $B$, $x_{q,Q}$,
and $m_{q,Q}$ are the parameters for the sum of the c.m. and gluon
fluctuation effects, bag pressure, lowest eigenvalues for the quarks
$q$ or $Q$, respectively, and the corresponding current quark masses.
$z_N$ and $B$ ($z_C$) are fixed by fitting the nucleon
(heavy baryon) mass in free space. For the current quark masses
we use $(m_{u,d},m_s,m_c,m_b) = (5,250,1300,4200)$ MeV,
where the values for $m_c$ and $m_b$ are  
the averaged values from Refs.~\cite{PDG1} and~\cite{PDG2}, respectively, 
and these values were used in 
Refs.~\cite{TK,TKhypbc,TsushimaD,Alexjpsi,Alexd}.
Then, we obtain the bag pressure $B = (170$ MeV$)^4$, by choosing the
bag radius for the nucleon in free space $R_N = 0.8$ fm.
The quark-meson coupling constants, which are determined so as 
to reproduce the saturation properties of symmetric nuclear matter are,  
($g^q_\sigma, g^q_\omega, g^q_\rho$) = ($5.69, 2.72, 9.33$), 
where $g_\sigma \equiv g^N_\sigma \equiv 
3 g^q_\sigma S_N(0) = 3 \times 5.69 \times 0.483 
= 8.23$~\cite{Saitof}. 
(See Eq.~(\ref{Csigma}) with the replacement, $C \to N$.)
The parameters $z_j$, and the bag radii $R_j$ in free space
$(j=N,\Lambda,\Sigma,\Xi,\Lambda_c^+,\Sigma_c,\Xi_c,\Lambda_b)$
and also some quantities calculated at normal nuclear mater density
$\rho_B = 0.15 fm^{-3}$, are listed in Table~\ref{bagparam},
together with the free space masses~\cite{PDG}.
(Nuclear mater limit will be discussed in Section~2.2.)

\begin{table}[htbp]
\begin{center}
\caption{The bag parameters,
the baryon masses and the bag radii in free space
[at normal nuclear matter density, $\rho_B = 0.15 fm^{-3}$]
$z_j, R_j$ and $M_j$ [$R_j^*$ and $M_j^\star$]
$(j=N,\Lambda,\Lambda^+_c,\Sigma_c,\Xi_c,\Lambda_b)$, respectively.
}
\label{bagparam}
\begin{tabular}[t]{c|ccc|cc}
\hline
j &$z_j$ &$M_j$ (MeV) &$R_j$ (fm) &$M_j^\star$ (MeV) &$R_j^*$ (fm)\\
\hline
$N$           &3.295 &939.0  &0.800 &754.5  &0.786\\
$\Lambda$     &3.131 &1115.7 &0.806 &992.7  &0.803\\
$\Sigma$      &2.810 &1193.1 &0.827 &1070.4 &0.824\\
$\Xi$         &2.860 &1318.1 &0.820 &1256.7 &0.818\\
$\Lambda^+_c$ &1.766 &2284.9 &0.846 &2162.5 &0.843\\
$\Sigma_c$    &1.033 &2452.0 &0.885 &2330.2 &0.882\\
$\Xi_c$       &1.564 &2469.1 &0.853 &2408.0 &0.851\\
$\Lambda_b$   &-0.643&5624.0 &0.930 &5502.9 &0.928\\
\end{tabular}
\end{center}
\end{table}
%

At the hadronic level, the entire information 
on the quark dynamics is condensed into the effective couplings  
$C_{N,C}(\sigma)$ of Eq.~(\ref{eqsigma}).
Furthermore, when $C_{N,C}(\sigma) = 1$, which corresponds to 
a structureless nucleon or a heavy baryon, the equations of motion
given by Eqs.~(\ref{eqdiracn})~-~(\ref{eqcoulomb})
can be identified with those derived 
from QHD~\cite{mar,ruf,Jennings}, 
except for the terms arising from the tensor coupling and the non-linear 
scalar and/or vector field interactions introduced beyond the naive QHD.

The explicit expressions for coupled differential
equations to obtain various fields and baryon wave
functions to describe a heavy baryon hypernuclear system, 
for the spherically symmetric nucleus plus one heavy baryon 
configuration 
can be obtained by replacing the corresponding indices 
from hyperon to heavy baryon ($Y \to C$) in Ref.~\cite{Tsushima_hyp}.

%
\subsection{Nuclear matter limit}

Here, we summarize the results and discussions given in 
Refs.~\cite{TK,Tsushima_hyp} for a limit of nuclear matter. 
New things in this subsection are some results 
given in Tables~\ref{bagparam} and~\ref{slope}, and 
to inform the similar treatments (parameterizations)  
also work for charmed and bottom baryons  
as will be shown in Eqs.~(\ref{cynsigma}) and~(\ref{Mstar}), 
which may not be so trivial.
In the nuclear matter limit all meson fields
become constant, and we denote the mean-values of the
$\omega$ and $\sigma$ fields by
$\overline{\omega}$ and $\overline{\sigma}$, respectively.
Then, equations for the $\overline{\omega}$ and
self-consistency condition for the $\overline{\sigma}$
are given by~\cite{Guichon,Guichonf,Saitof,matter} 
\bg
\overline{\omega} &=& \frac{4}{(2\pi)^3}\int d^3 k\, \theta (k_F - k)
= \frac{g_\omega}{m_\omega^2} \frac{2k_F^3}{3\pi^2}
= \frac{g_\omega}{m_\omega^2} \rho_B,
\label{omega_bar}\\
\overline{\sigma}&=&\frac{g_\sigma }{m_\sigma^2}C_N(\overline{\sigma})
\frac{4}{(2\pi)^3}\int d^3 k\, \theta (k_F - k)
\frac{M_N^{\star}(\overline{\sigma})}
{\sqrt{M_N^{\star 2}(\overline{\sigma})+\vec{k}^2}}
= \frac{g_\sigma }{m_\sigma^2}C_N(\overline{\sigma}) \rho_s,
\label{scc}
\en
where $g_{\sigma} = 3 g^q_\sigma S_N(0)$ 
(see Eq.~(\ref{Csigma}) by the replacement, $C \to N$), $k_F$ is the
Fermi momentum, $\rho_B$ and $\rho_s$ are the baryon and scalar
densities, respectively.
Note that $M_N^\star (\overline{\sigma})$   
in Eq.~(\ref{scc}), must be calculated 
self-consistently by the MIT bag model, through 
Eqs.~(\ref{gsigma}) - (\ref{diracu}).
This self-consistency equation for the $\overline{\sigma}$
is the same as that in QHD, except that in the latter one has
$C_N(\overline{\sigma})=1$~\cite{QHD}. 
Using the mean field value $\overline{\sigma}$, 
the corresponding quantity for the heavy baryon $C$, 
$C_C(\overline{\sigma})$, can be also calculated using 
Eqs.~(\ref{gsigma})~-~(\ref{diracQ}) neglecting 
the effect of a single heavy baryon on the
mean field value $\overline{\sigma}$, in infinite nuclear matter.

It has been found that the
function $C_j(\overline{\sigma}) 
(j=N,\Lambda,\Sigma,\Xi,\Lambda_c^+,\Sigma_c,\Xi_c,\Lambda_b)$
can be parametrized as a linear
form in the $\sigma$ field, $g_{\sigma}\overline{\sigma}$, for practical
calculations~\cite{Tsushima_hyp,Guichonf,Saitof}: 
\begin{equation}
C_j (\overline{\sigma}) = 1 - a_j 
\times (g_{\sigma} \overline{\sigma}),\hspace{1em} 
(j=N,\Lambda,\Sigma,\Xi,\Lambda_c^+,\Sigma_c,\Xi_c,\Lambda_b). 
\label{cynsigma}
\end{equation}
The values obtained for $a_j$ are listed in Table~\ref{slope}.
This parameterization works very well up to
about three times of normal nuclear matter density $\rho_B \simeq 3 \rho_0$  
($\rho_0 \simeq 0.15$ fm$^{-3}$).
Then, the effective masses for the baryon, $j$, in nuclear matter
are well approximated by~\cite{Tsushima_hyp,Guichonf,Saitof}
\bge
M^\star_j \simeq M_j - \frac{n_q}{3} g_\sigma
\left[1-\frac{a_j}{2}(g_\sigma\overline{\sigma})\right]\overline{\sigma},
\hspace{2ex}(j=N,\Lambda,\Sigma,\Xi,\Lambda_c^+,\Sigma_c,\Xi_c,\Lambda_b),
\label{Mstar}
\ene
with $n_q$ being the number of light quarks in the baryon $j$.
For the field strength $g_\sigma \overline{\sigma}$
versus baryon density, one can find in Ref.~\cite{Guichonf}.
%
\begin{table}[htbp]
\begin{center}
\caption{The slope parameters, $a_j\,\,
(j=N,\Lambda,\Sigma,\Xi,\Lambda_c^+,\Sigma_c,\Xi_c,\Lambda_b)$.}
\label{slope}
\begin{tabular}[t]{c|c||c|c}
\hline
$a_j$ &$\times 10^{-4}$ MeV$^{-1}$ &$a_j$ &$\times 10^{-4}$ MeV$^{-1}$ \\
\hline
$a_N$           &8.8  &$a_{\Lambda_b}$   &10.9 \\
$a_\Lambda$     &9.3  &$a_{\Lambda_c^+}$ &9.8 \\
$a_{\Sigma}$    &9.5  &$a_{\Sigma_c}$    &10.3 \\
$a_{\Xi}$       &9.4  &$a_{\Xi_c}$       &9.9 \\
\end{tabular}
\end{center}
\end{table}
%

\begin{figure}[hbt]
\begin{center}
\epsfig{file=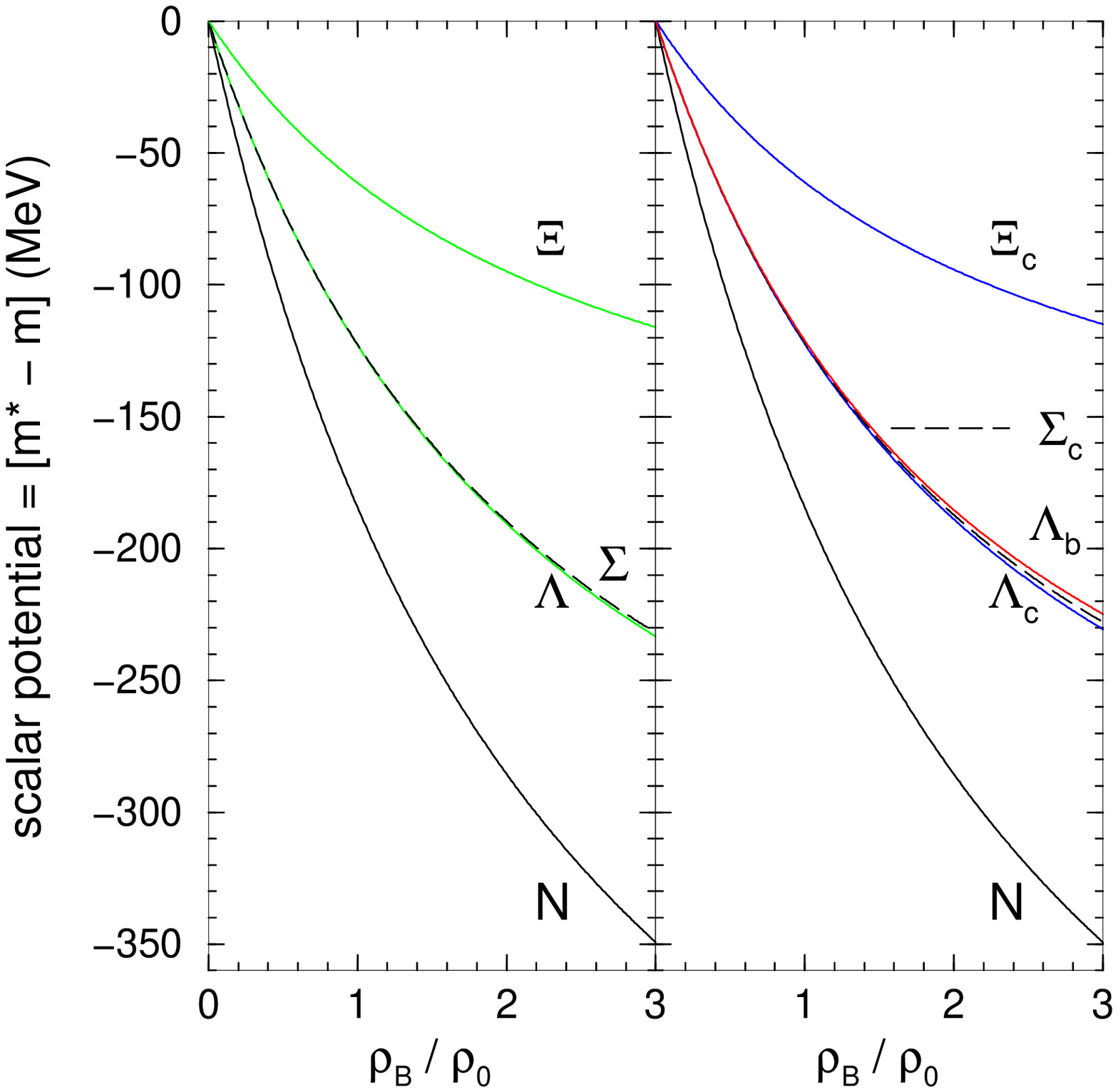,height=14cm,angle=-90}
\caption{
Scalar potentials for hyperons, low-lying charmed and bottom
baryons in symmetric nuclear matter taken from Ref.~\protect\cite{TK}.
\label{spotential}
}
\end{center}
\end{figure}

Next, we discuss briefly the 
scalar and vector potentials of charmed and bottom 
baryons in nuclear matter~\cite{TK}.
The scalar ($V^{j}_s$) and vector ($V^{j}_v$) potentials 
for the baryon $j$, in nuclear matter are given by 
\bg
V^j_s &=& m^\star_j - m_j,\,\,\label{spot}
\\
V^j_v &=&
  n_q {V}^q_\omega + I^j_3 V^q_\rho,
\label{vpot}
\en
where $I^j_3$ is the third component of isospin projection   
of the baryon $j$. Thus, the vector potential for a heavy baryon
with charm or bottom quark(s), is equal to that of the hyperon with
the same light quark number in the QMC model.
Calculated results for the scalar potentials for the baryons 
in symmetric nuclear matter are shown in Fig.~\ref{spotential}, 
which is taken from Ref.~\cite{TK} by omitting those for mesons. 
From the results it is confirmed that
the scalar potential for the baryon $j$, $V_s^j$,
follows a simple light quark number counting rule:
\bge
V_s^j \simeq \frac{n_q}{3} V_s^N,
\ene
where $n_q$ is the number of light quarks in
the baryon $j$, and $V_s^N$ is the scalar potential for the nucleon.
(See Eq.(\ref{spot}).)
It is interesting to note that, the baryons with charm and bottom quarks 
($\Xi_c$ is a quark flavor configuration, $qsc$), 
show features very similar
to those of the corresponding strange hyperons with the same 
light quark numbers.
Then, ignoring the Coulomb force although it is important in a 
realistic hypernucleus, we may expect that these heavy 
baryons with charm or bottom quarks, will also form some 
heavy baryon hypernuclei at this stage, as the
strange hyperons do. (Recall that the repulsive, vector
potentials are the same for the corresponding hyperons with the same light
quark numbers.)

%
\section{Calculations}\label{Calculation}

In this section, we discuss parameters used in the calculations:   
spin-orbit force in the QMC model, 
the Pauli blocking effect at the quark level, 
and the effect of channel coupling, $\Sigma_{c,b} N - \Lambda_{c,b} N$.
The parameters are the same as those used in the study of 
strange hypernuclei~\cite{Tsushima_hyp}, except for those 
related to charmed and bottom hypernuclei given in Section~2,  
but the treatment is the same as that of Ref.~\cite{Tsushima_hyp}.
More detailed discussions may be found in Ref.~\cite{Tsushima_hyp}.

\subsection{Spin-orbit force in the QMC model}\label{sopot}
The origin of the spin orbit force for a composite nucleon moving
through scalar and vector fields which vary with position is explained
in detail in Ref.~\cite{Guichonf}. 
For the $\Lambda, \Sigma$ and $\Xi$ cases are 
explained in Ref.~\cite{Tsushima_hyp}.
Although the spin-orbit splittings for the nucleon calculated
in QMC are already somewhat smaller~\cite{Guichonf,Saitof}, 
it was demonstrated that much 
smaller spin-orbit splittings were obtained 
for the $\Lambda$ in QMC~\cite{Tsushima_hyp}. 
These will turn out to be even smaller 
for the heavy baryon hypernuclei.

In order to include the spin-orbit potential approximately,  
e.g., for the $\Lambda^+_c$, 
we add perturbatively the correction due to the vector potential,
$ -\frac{2}{2 M^{\star 2}_{\Lambda^+_c} (\vec{r}) r}
\, \left( \frac{d}{dr} g^{\Lambda^+_c}_\omega \omega(\vec{r}) \right)
\vec{l}\cdot\vec{s}$,
to the single-particle energies obtained with the Dirac
equation, in the same way as that added for the strange 
hypernuclei~\cite{Tsushima_hyp}.
This may correspond to a correct spin-orbit force which
is calculated by the underlying quark
model~\cite{Tsushima_hyp,Guichonf}:
\begin{equation}
V^{\Lambda^+_c}_{S.O.}(\vec{r}) \vec{l}\cdot\vec{s}
= - \frac{1}{2 M^{\star 2}_{\Lambda^+_c} (\vec{r}) r}
\, \left( \frac{d}{dr} [ M^\star_{\Lambda^+_c} (\vec{r})
+ g^{\Lambda^+_c}_\omega \omega(\vec{r}) ] \right) \vec{l}\cdot\vec{s},
\label{soQMC}
\end{equation}
since the Dirac equation at the hadronic level solved in usual QHD-type
models leads to:
\begin{equation}
V^{\Lambda^+_c}_{S.O.}(\vec{r}) \vec{l}\cdot\vec{s}
= - \frac{1}{2 M^{\star 2}_{\Lambda^+_c} (\vec{r}) r}
\, \left( \frac{d}{dr} [ M^\star_{\Lambda^+_c} (\vec{r})
- g^{\Lambda^+_c}_\omega \omega(\vec{r}) ] \right) \vec{l}\cdot\vec{s},
\label{soQHD}
\end{equation}
which has the opposite sign for the vector potential,
$g^{\Lambda^+_c}_\omega \omega(\vec{r})$.
The correction to the spin-orbit force in the QMC model, 
appears due to the structure of the baryon (finite size).  
This may also be modeled at the hadronic level of the Dirac equation by
adding a tensor interaction, motivated by the quark
model~\cite{Jennings2,Cohen}.
In addition, one boson exchange model with underlying approximate 
SU(3) symmetry in strong interactions, also leads to a weaker
spin-orbit forces for the (strange) hyperon-nucleon ($YN$)
than that for the nucleon-nucleon ($NN$)~\cite{Gal}.

In practice, because of its heavy mass
($M^\star_{\Lambda^+_c}$), contribution to the single-particle energies
from the spin-orbit potential both with or without
the inclusion of the correction term, turns
out to be even smaller than that for the
$\Lambda$ hypernuclei, and further smaller for 
the $\Lambda_b$ hypernuclei~\cite{TKhypbc}.
Contribution from the spin-orbit potential with the correction term
is typically of order $0.01$ MeV,
and even the largest case is $\simeq 0.5$ MeV, among all the heavy 
baryon hypernuclei calculated in this study.
This can be understood when one considers the limit,
$M^\star_{\Lambda^+_c} \to \infty$
in Eq.~(\ref{soQMC}), where the quantity inside the square brackets varies
smoothly from an order of hundred MeV to zero near the surface of
the hypernucleus, and the derivative with respect to $r$ is finite.
The results for the other heavy baryon hypernuclei are quite similar, 
and spin-orbit forces give only tiny contributions for the single-particle
energies. Thus, the spin-orbit splittings are expected to be small.

%
\subsection{Effects of the Pauli blocking and
channel coupling}\label{Pauli}

Below, we briefly explain the effects of the Pauli blocking 
due to the quark structure, and the channel 
coupling~\cite{Tsushima_hyp}.

Although the present treatment is based on the underlying quark structure of 
the nucleon and heavy baryons, it is missing 
an explicit inclusion of the Pauli blocking effect  
at the quark level among the $u$ and $d$ quarks in the core nucleons and 
the heavy baryons. For the strange hypernuclei,
the effect of the Pauli blocking 
was included in a specific way at the hadronic level.
We follow the treatment of Ref.~\cite{Tsushima_hyp}, and include 
effective Pauli blocking and also the $\Sigma_{c,b} N - \Lambda_{c,b} N$
channel coupling effects, although the channel coupling 
effect is expected to be smaller than that for the 
$\Sigma N - \Lambda N$, since the mass difference 
for the former is larger than the latter.
Thus, the channel coupling effect that is 
included for the $\Sigma_c$ hypernuclei in the present study
should be regarded as the limiting case.

Following Ref.~\cite{Tsushima_hyp}, we assume  
that the Pauli blocking effect is 
simply proportional to the nucleon baryonic density 
(or $u$ and $d$ total light quark number density), 
although one could consider a more complicated density dependence.
Then, the Dirac equation for the heavy baryon C, Eq.~(\ref{eqdiracy}),  
may be modified by 
\bge
[i\gamma \cdot \partial - M^{\star}_C(\sigma)-
(\, \lambda_C \rho_B(\vec{r}) + g^C_\omega \omega(\vec{r}) 
+ g_\rho I^C_3 b(\vec{r})
+ e Q_C A(\vec{r}) \,)
\gamma_0 ] \psi_C(\vec{r}) = 0, \label{mdiracy}
\ene
where, $\rho_B(\vec{r})$ is the baryonic density at the position $\vec{r}$ 
in the heavy baryon hypernucleus due to the core nucleons, and $\lambda_C$ is a 
constant, taken the same value to that 
determined for the strange hyperon. We note that the Pauli blocking 
effect associated with the light quarks in the heavy baryon should also 
lead to some repulsion for the nucleons. 
Contrary to the comment made in Ref.~\cite{Tsushima_hyp}, 
the actual calculations in this study 
(and in Ref.~\cite{Tsushima_hyp}) 
include the effect of the modification made in Eq.~(\ref{mdiracy}) 
in a self-consistent manner.
In practice, the above modified Eq.~(\ref{mdiracy}) has been inserted  
again into a system of coupled differential equations for the spherical 
static (heavy baryon) hypernuclear system, from Eq.~(\ref{eqdiracn}) to 
Eq.~(\ref{energy}), and calculations are performed again.
Thus, the response of the core nucleons due to the modification 
in Eq.~(\ref{mdiracy}) for the heavy baryon, is included.

For the $\Lambda$ hypernuclei~\cite{Tsushima_hyp} the parameter corresponding 
to $\lambda_C$ in Eq.~(\ref{mdiracy}) was chosen   
to reproduce the empirical single particle energy for the 
$1s_{1/2}$ in $^{209}_\Lambda$Pb, -27.0 MeV~\cite{aji}.
The fitted value for the constant was, 
$\lambda_\Lambda = 60.25$ MeV (fm)$^3$.
Thus, we use the same value also for the $\Lambda_{c,b}$, i.e.,  
$\lambda_{\Lambda_{c,b}} = 60.25$ MeV (fm)$^3$.
For the $\Sigma_c$ and $\Xi_c$ hypernuclei, we can deduce 
the constants $\lambda_{\Sigma_c,\Xi_c}$, 
corresponding to the effective Pauli blocking effect as,  
$\lambda_{\Sigma_c} = \lambda_{\Lambda_c}$, 
and $\lambda_{\Xi_c} = \frac{1}{2} \lambda_{\Lambda_c}$,  
by counting the total number of $u$ and $d$ quarks in the heavy baryons.

One may regard that this fitted value   
includes also the attractive 
$\Lambda_{c,b} N \rightarrow \Sigma_{c,b} N$
channel coupling effect for the $\Lambda_{c,b}$ single-particle energies, 
since the value fitted for the $\Lambda$ hypernuclei 
includes the experimentally observed effect of the 
$\Lambda N \rightarrow \Sigma N$ channel coupling.
As for the $\Sigma_c N \rightarrow \Lambda_c N$ channel 
coupling effect, it must be included 
in addition    
to reproduce the relative repulsive energy shift in the 
single-particle energies for the $\Sigma_c$ hypernuclei, 
as was done for the $\Sigma$ hypernuclei.

The $\Sigma N - \Lambda N$ channel coupling was estimated 
using the Nijmegen potential in Ref.~\cite{Tsushima_hyp}. 
The effect for the $\Sigma$ 
was included by assuming the same form as that 
for the effective Pauli blocking via $\lambda_\Sigma \rho_B(r)$, 
and adjusted the parameter, $\lambda_\Sigma = \lambda_\Lambda \to 
\tilde{\lambda}_\Sigma \neq \lambda_\Sigma$, to reproduce 
this difference in the 
single-particle energy for the $1s_{1/2}$ in $^{\Sigma^0}_{209}$Pb, 
namely, $-19.6 = -26.9 (\simeq -27.0) + 7.3$ MeV. 
The value obtained for $\tilde{\lambda}_\Sigma$ in this way was, 
$\tilde{\lambda}_\Sigma$ = 110.6 MeV (fm)$^3$.
Thus, we use the same value, 
$\tilde{\lambda}_{\Sigma_c}$ = 110.6 MeV (fm)$^3$, although  
the effect is expected to be smaller considering the mass differences, 
$m_\Sigma - m_\Lambda \simeq 77$ MeV and 
$m_{\Sigma_c} - m_{\Lambda_c^+} \simeq 166$ MeV~\cite{PDG}.
As for the $\Xi N - \Lambda \Lambda$ channel coupling
the effect was estimated to be negligible in Ref.~\cite{Tsushima_hyp}, 
and we also neglect the effect of channel coupling on 
$\Xi_c$ in this study.

\section{Results for heavy baryon hypernuclei}
%

%
Before presenting the results, we summarize in 
Table~\ref{parameter} the parameters at the hadronic level 
fixed by the saturation properties of
symmetric nuclear matter. 
Concerning the parameters for the $\sigma$ field, we note that
the properties of nuclear matter only fix the ratio, ($g_\sigma/m_\sigma$),
with a chosen value, $m_\sigma = 550$ MeV.
Keeping this ratio to be a constant,
the value $m_\sigma = 418$ MeV for finite nuclei (hypernuclei)
is obtained by fitting the r.m.s. charge radius of $^{40}$Ca
to the experimental value,
$r_{{\rm ch}}(^{40}$Ca$)$ = 3.48 fm~\cite{Saitof}.
(See also Section 2.1 for the value of $g_\sigma$ in nuclear matter.)

\begin{table}[htbp]
\begin{center}
\caption{Parameters at the hadronic
level for finite nuclei~\protect\cite{Saitof}.}
\label{parameter}
\begin{tabular}[t]{c|cc}
\hline
field &mass (MeV) &$g^2/4\pi\, (e^2/4\pi)$\\
\hline
$\sigma$ &418 &3.12\\
$\omega$ &783 &5.31\\
$\rho$   &770 &6.93\\
$A$      &0   &1/137.036\\
\end{tabular}
\end{center}
\end{table}
%

First, in Tables~\ref{spe1} and~\ref{spe2}, we 
summarize the calculated 
single-particle energies for $^{17}_j$O, $^{41}_j$Ca, 
$^{49}_j$Ca, $^{91}_j$Zr and $^{209}_j$Pb 
($j=\Lambda,\Lambda_c^+,\Sigma_c^{0,+,++},\Xi_c^{0,+},\Lambda_b$) 
hypernuclei, together with the experimental data~\cite{chr,aji} 
for the $\Lambda$ hypernuclei. 
The results for the $\Lambda_c^+$ and $\Lambda_b$, and $\Lambda$ 
hypernuclei are taken from 
Refs.~\cite{TKhypbc}, and~\cite{Tsushima_hyp}, respectively 
for comparison.
In the calculation, 
we have searched for the single-particle states up to the same highest 
state as that of the core neutrons in each hypernucleus, since the 
deeper levels are usually easier to observe in experiment.
%
\begin{table}[htbp]
\begin{center}
\caption{Single-particle energies (in MeV)
for $^{17}_j$O, $^{41}_j$Ca and $^{49}_j$Ca  
($j = \Lambda, \Lambda^+_c, \Sigma_c, \Xi_c, \Lambda_b$).
Results for $\Lambda_c^+$ and $\Lambda_b$, and $\Lambda$ 
hypernuclei are taken from Refs.~\protect\cite{TKhypbc}, 
and~\cite{Tsushima_hyp}, respectively.
Experimental data are taken from Ref.~\protect\cite{chr}.
Spin-orbit splittings for the $\Lambda$ hypernuclei 
are not well determined by the experiments.} 
\label{spe1}
\begin{tabular}[t]{c|ccccccccc}
\hline \hline
&$^{16}_\Lambda$O (Expt.)
&$^{17}_\Lambda$O    &$^{17}_{\Lambda^+_c}$O
&$^{17}_{\Sigma^0_c}$O &$^{17}_{\Sigma^+_c}$O &$^{17}_{\Sigma^{++}_c}$O
&$^{17}_{\Xi^0_c}$O    &$^{17}_{\Xi^+_c}$O &$^{17}_{\Lambda_b}$O\\
\hline \hline
$1s_{1/2}$&-12.5      &-14.1&-12.8&---  &-7.5&---&-7.9&-2.1&-19.6\\
$1p_{3/2}$&-2.5 ($1p$)&-5.1 &-7.3 &-10.7&-4.0&---&-3.5&--- &-16.5\\
$1p_{1/2}$&-2.5 ($1p$)&-5.0 &-7.3 &-10.2&-3.6&---&-3.5&--- &-16.5\\ \\
\hline \hline
&$^{40}_\Lambda$Ca (Expt.)
&$^{41}_\Lambda$Ca    &$^{41}_{\Lambda^+_c}$Ca
&$^{41}_{\Sigma^0_c}$Ca &$^{41}_{\Sigma^+_c}$Ca &$^{41}_{\Sigma^{++}_c}$Ca
&$^{41}_{\Xi^0_c}$Ca    &$^{41}_{\Xi^+_c}$Ca &$^{41}_{\Lambda_b}$Ca\\
\hline \hline
$1s_{1/2}$&-20.0       &-19.5&-12.8&-16.3&-6.3&---&-9.9&-1.0&-23.0\\
$1p_{3/2}$&-12.0 ($1p$)&-12.3&-9.2 &-13.2&-4.1&---&-6.7&--- &-20.9\\
$1p_{1/2}$&-12.0 ($1p$)&-12.3&-9.1 &-12.9&-3.8&---&-6.7&--- &-20.9\\
$1d_{5/2}$&            &-4.7 &-4.8 &-9.9 &--- &---&-3.3&--- &-18.4\\
$2s_{1/2}$&            &-3.5 &-3.4 &-9.3 &--- &---&-2.8&--- &-17.4\\
$1d_{3/2}$&            &-4.6 &-4.8 &-9.4 &--- &---&-3.3&--- &-18.4\\ \\
\hline \hline
&--- 
&$^{49}_\Lambda$Ca    &$^{49}_{\Lambda^+_c}$Ca
&$^{49}_{\Sigma^0_c}$Ca &$^{49}_{\Sigma^+_c}$Ca &$^{49}_{\Sigma^{++}_c}$Ca
&$^{49}_{\Xi_c^0}$Ca    &$^{49}_{\Xi_c^+}$Ca &$^{49}_{\Lambda_b}$Ca\\ 
\hline \hline
$1s_{1/2}$&           &-21.0&-14.3&--- &-7.4&-5.0&-7.8&-5.1&-24.4\\
$1p_{3/2}$&           &-13.9&-10.6&-7.2&-5.1&-3.4&-4.1&-2.8&-22.2\\
$1p_{1/2}$&           &-13.8&-10.6&-7.0&-4.9&-3.2&-4.1&-2.8&-22.2\\
$1d_{5/2}$&           &-6.5 &-6.5 &-4.0&-2.3&--- &-0.9&--- &-19.5\\
$2s_{1/2}$&           &-5.4 &-5.3 &-4.9&--- &--- &-1.5&--- &-18.8\\
$1d_{3/2}$&           &-6.4 &-6.4 &-3.6&-1.9&--- &-1.0&--- &-19.5\\
$1f_{7/2}$&           &---  &-2.0 &--- &--- &--- &--- &--- &-16.8\\
\end{tabular}
\end{center}
\end{table}
%
%

%
\begin{table}[htbp]
\begin{center}
\caption{Single-particle energies (in MeV)
for $^{91}_j$Zr and $^{208}_j$Pb  
($j = \Lambda, \Lambda^+_c, \Sigma_c, \Xi_c, \Lambda_b$).
Results for $\Lambda_c^+$ and $\Lambda_b$, and $\Lambda$
hypernuclei are taken from Refs.~\protect\cite{TKhypbc},
and~\cite{Tsushima_hyp}, respectively.
Experimental data are taken from Ref.~\protect\cite{aji}.
Spin-orbit splittings for the $\Lambda$ hypernuclei 
are not well determined by the experiments. 
The values given inside the brackets for $^{91}_{\Sigma^0_c}$Zr are those 
obtained by switching off both the Pauli blocking and the channel 
coupling effects simultaneously by setting $\tilde{\lambda}_{\Sigma_c} = 0$.
}
\label{spe2}
\begin{tabular}[t]{c|ccccccccc}
\hline \hline
&$^{89}_\Lambda$Yb (Expt.)
&$^{91}_\Lambda$Zr    &$^{91}_{\Lambda^+_c}$Zr
&$^{91}_{\Sigma^0_c}$Zr &$^{91}_{\Sigma^+_c}$Zr &$^{91}_{\Sigma^{++}_c}$Zr
&$^{91}_{\Xi^0_c}$Zr    &$^{91}_{\Xi^+_c}$Zr &$^{91}_{\Lambda_b}$Zr\\ 
\hline \hline
$1s_{1/2}$&-22.5       &-23.9&-10.8&--- (---)  &-3.7 &---&-9.3&---&-25.7\\
$1p_{3/2}$&-16.0 ($1p$)&-18.4&-8.7 &-10.2 (-26.5) &-2.3 &---&-6.6&---&-24.2\\
$1p_{1/2}$&-16.0 ($1p$)&-18.4&-8.7 &-10.1 (-26.4) &-2.1 &---&-6.7&---&-24.2\\
$1d_{5/2}$&-9.0  ($1d$)&-12.3&-5.8 &-7.6  (-21.8) &---  &---&-4.0&---&-22.4\\
$2s_{1/2}$&            &-10.8&-3.9 &-8.1  (-23.0) &---  &---&-3.9&---&-21.6\\
$1d_{3/2}$&-9.0  ($1d$)&-12.3&-5.8 &-7.3  (-21.6)&---  &---&-4.0&---&-22.4\\
$1f_{7/2}$&-2.0  ($1f$)&-5.9 &-2.4 &-5.1  (-17.1) &---  &---&-1.3&---&-20.4\\
$2p_{3/2}$&            &-4.2 &---  &-5.0  (-16.5) &---  &---&-1.3&---&-19.5\\
$1f_{5/2}$&-2.0  ($1f$)&-5.8 &-2.4 &-4.7  (-16.8) &---  &---&-1.4&---&-20.4\\
$2p_{1/2}$&            &-4.1 &---  &-4.9  (-16.3) &---  &---&-1.3&---&-19.5\\
$1g_{9/2}$&            &---  &---  &-2.4  (-12.4) &---  &---&--- &---&-18.1\\ \\
\hline \hline
&$^{208}_\Lambda$Pb (Expt.)
&$^{209}_\Lambda$Pb    &$^{209}_{\Lambda^+_c}$Pb
&$^{209}_{\Sigma^0_c}$Pb &$^{209}_{\Sigma^+_c}$Pb &$^{209}_{\Sigma^{++}_c}$Pb
&$^{209}_{\Xi^0_c}$Pb    &$^{209}_{\Xi^+_c}$Pb &$^{209}_{\Lambda_b}$Pb\\ 
\hline \hline
$1s_{1/2}$&-27.0       &-27.0&-5.2&-7.5&---&---&-6.7&---&-27.4\\
$1p_{3/2}$&-22.0 ($1p$)&-23.4&-4.1&-6.6&---&---&-5.4&---&-26.6\\
$1p_{1/2}$&-22.0 ($1p$)&-23.4&-4.0&-6.5&---&---&-5.5&---&-26.6\\
$1d_{5/2}$&-17.0 ($1d$)&-19.1&-2.4&-5.3&---&---&-3.9&---&-25.4\\
$2s_{1/2}$&            &-17.6&--- &--- &---&---&-3.3&---&-24.7\\
$1d_{3/2}$&-17.0 ($1d$)&-19.1&-2.4&-5.1&---&---&-4.0&---&-25.4\\
$1f_{7/2}$&-12.0 ($1f$)&-14.4&--- &-3.8&---&---&-2.2&---&-24.1\\
$2p_{3/2}$&            &-12.4&--- &--- &---&---&-1.4&---&-23.2\\
$1f_{5/2}$&-12.0 ($1f$)&-14.3&--- &-3.5&---&---&-2.3&---&-24.1\\
$2p_{1/2}$&            &-12.4&--- &--- &---&---&-1.5&---&-23.2\\
$1g_{9/2}$&-7.0  ($1g$)&-9.3 &--- &-2.1&---&---&--- &---&-22.6\\
$1g_{7/2}$&-7.0  ($1g$)&-9.2 &--- &-1.8&---&---&--- &---&-22.6\\
$1h_{11/2}$&           &-3.9 &--- &--- &---&---&--- &---&-21.0\\
$2d_{5/2}$&            &-7.0 &--- &--- &---&---&--- &---&-21.7\\
$2d_{3/2}$&            &-7.0 &--- &--- &---&---&--- &---&-21.7\\
$1h_{9/2}$&            &-3.8 &--- &--- &---&---&--- &---&-21.0\\
$3s_{1/2}$&            &-6.1 &--- &--- &---&---&--- &---&-21.3\\
$2f_{7/2}$&            &-1.7 &--- &--- &---&---&--- &---&-20.1\\
$3p_{3/2}$&            &-1.0 &--- &--- &---&---&--- &---&-19.6\\
$2f_{5/2}$&            &-1.7 &--- &--- &---&---&--- &---&-20.1\\
$3p_{1/2}$&            &-1.0 &--- &--- &---&---&--- &---&-19.6\\
$1i_{13/2}$&           &---  &--- &--- &---&---&--- &---&-19.3\\
\end{tabular}
\end{center}
\end{table}
%
%

After a first look at the results shown in Tables~\ref{spe1} and~\ref{spe2} 
we notice: 
\begin{enumerate}
\item $\Sigma_c^{++}$ and $\Xi_c^+$ hypernuclei are very unlikely to be 
formed in usual circumstances.
Solid conclusions may not be drawn about $^{49}_{\Sigma_c^{++},\Xi_c^+}$Ca. 
Although results imply the formations of these hypernuclei, 
those numbers may be regarded within the uncertainties of the model and 
the approximations made in the calculation. 
Only a major alteration of the parameters may yield 
such bound hypernuclei.
\item $\Sigma_c^0$ and $\Sigma_c^+$ hypernuclei may have some 
possibilities to be formed.
However, due to a peculiar feature that the correct 
$1s_{1/2}$ state is not found in 
$^{17}_{\Sigma^0_c}$O, $^{49}_{\Sigma^0_c}$Ca and 
$^{91}_{\Sigma^0_c}$Zr, one needs some caution to take this 
statement naively.
(Discussions on this matter will be made later in detail.)
A calculational modification could increase confidence in 
predicting such hypernuclei.
\item $\Lambda_c^+, \Xi_c^0$ and $\Lambda_b$ hypernuclei 
are expected to be formed quite likely in a realistic situation.
However, for the $\Lambda_b$ hypernuclei, it will be very 
difficult to achieve such a resolution to distinguish 
the states experimentally. 
The question of experimental observation is not tackled realistically.
\item The Coulomb force plays a crucial role in forming (unforming) 
hypernuclei, by comparing the single-particle energies among/between 
the members within each multiplet, 
($\Lambda,\Lambda_c^+,\Lambda_b$), ($\Sigma_c^0,\Sigma_c^+,\Sigma_c^{++}$), 
and ($\Xi_c^0,\Xi_c^+$) hypernuclei.
\end{enumerate}
 
We discuss below in detail  
the $\Sigma_c^0$ hypernuclei single-particle energies,  
which have a peculiar feature. In some of the hypernuclei 
calculated, the $1s_{1/2}$ state could not be found in the following sense.
We have searched for the $1s_{1/2}$ state by varying matching point 
to get the eigenenergy and the corresponding eigen wave function for 
the $\Sigma_c^0$ in steps 0.1 fm, from 0.2 fm to 5.0 fm 
from the center of the hypernucleus. 
However, always a proper level with lower energy than those of the 
$1p_{1/2}, 1p_{3/2}$ and $2s_{1/2}$ states 
could not be found, although a smaller matching point than that  
for the $2s_{1/2}$ state and the Dirac's $\kappa = -1$
necessary to obtain the correct $1s_{1/2}$ wave function were input. 
In such a case, the single-particle energy 
level became equal or higher than that of 
the $2s_{1/2}$ state, which is also above the levels of 
the $1p_{1/2}$ and $1p_{3/2}$ states.  
In the search procedure, we always checked that the core 
nucleon single-particle energy levels are in correct order. 
But, whenever we get the lower single-particle energy level for the 
$1s_{1/2}$ state than those for the $1p_{1/2}, 1p_{3/2}$ 
and $2s_{1/2}$ states, the core neutron $1s_{1/2}$ energy level
jumped to become equal or larger than that of the $2s_{1/2}$,  
and we discarded the results in such cases. In this manner, the $1s_{1/2}$
state in $^{17}_{\Sigma_c^0}$O, $^{49}_{\Sigma_c^0}$Ca and 
$^{91}_{\Sigma_c^0}$Zr could not be found.

\begin{figure}[hbt]
\begin{center}
\epsfig{file=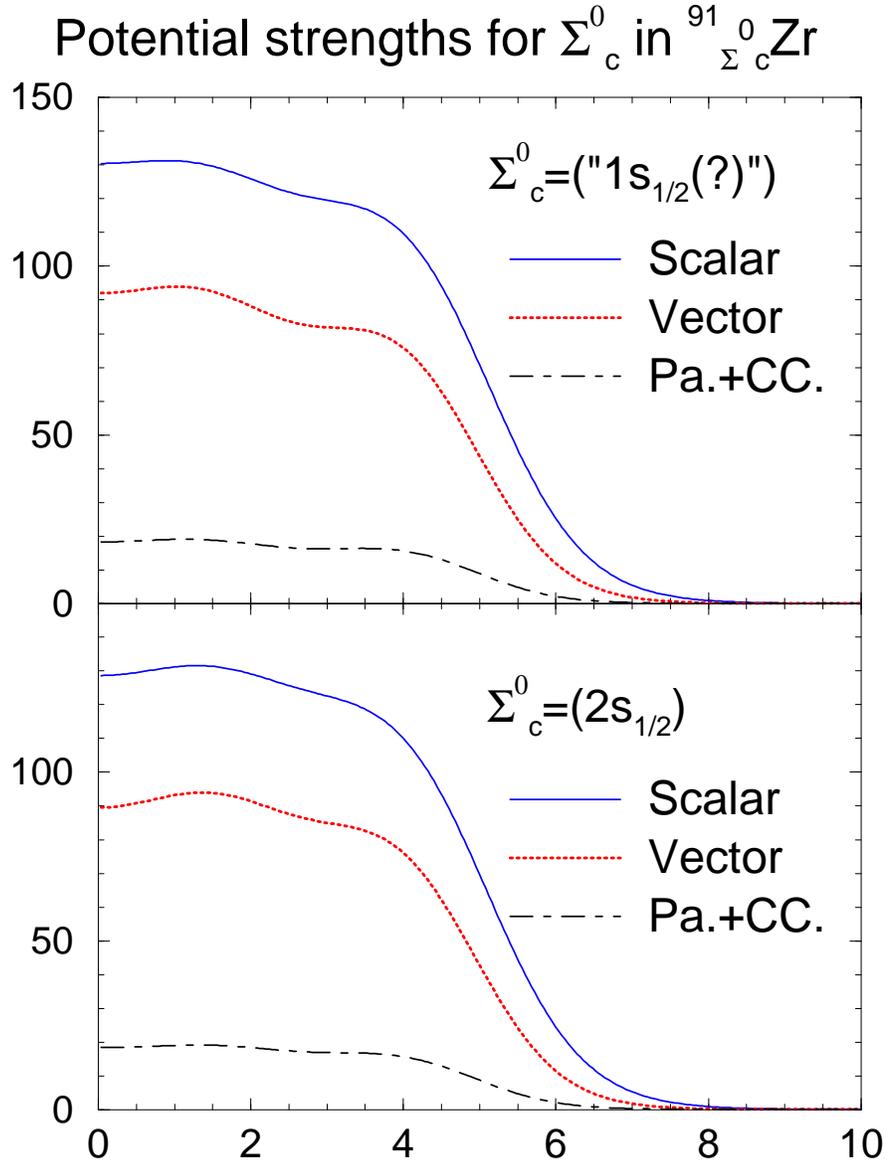,height=12cm,angle=-90}
\caption{Potential strengths for $\Sigma_c^{0,+}$ in 
$^{91}_{\Sigma_c^0}$Zr.
"Pa.+CC." stands for potential for
the effective Pauli blocking plus the channel coupling effects.
See the text for the meaning of "$1s_{1/2}(?)$".
\label{Sigc0zrpot}
}
\end{center}
\end{figure}

\begin{figure}[hbt]
\begin{center}
\epsfig{file=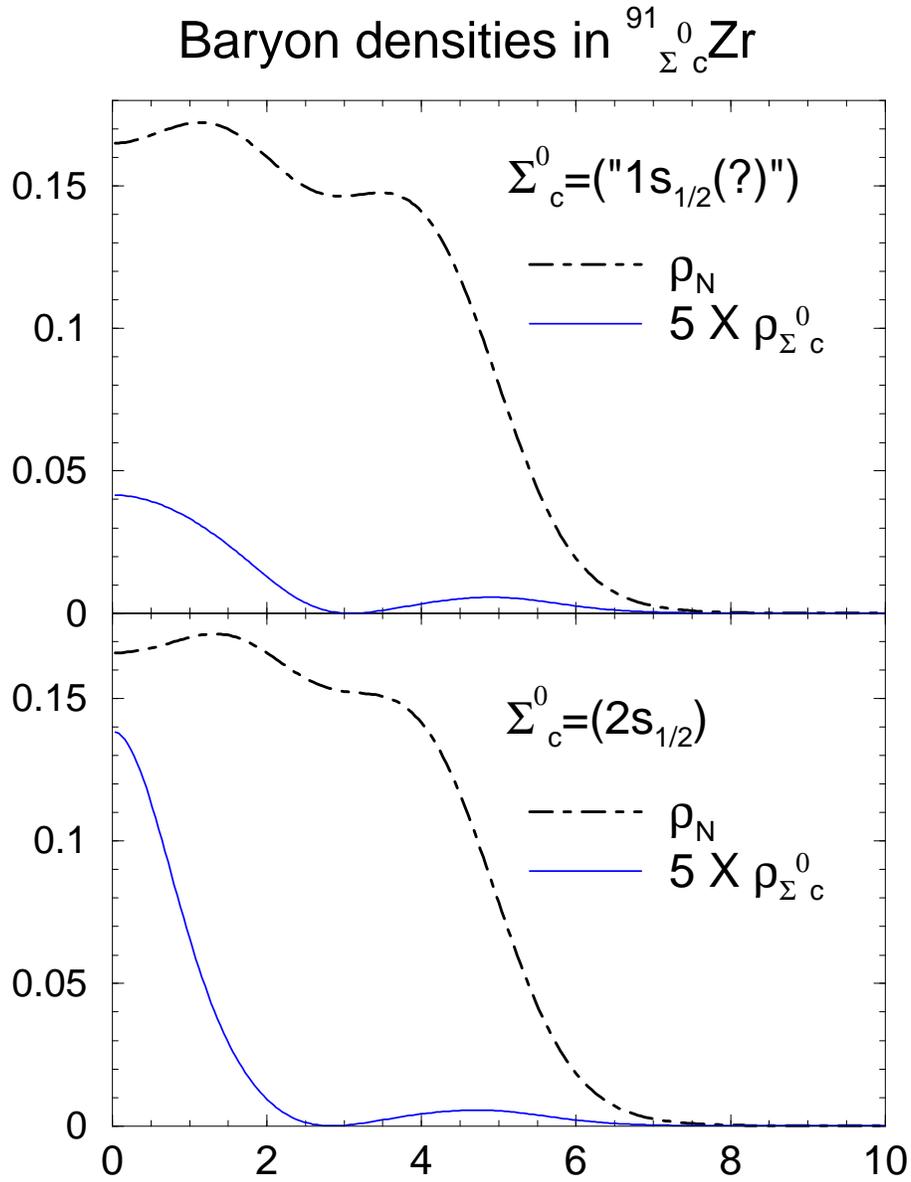,height=12cm,angle=-90}
\caption{Baryon density distributions for the 
$\Sigma_c^0$ multiplied by a factor 5, and nucleons  
in $^{91}_{\Sigma_c^0}$Zr.
\label{Sigc0zrbden}
}
\end{center}
\end{figure}

To get greater insight as to what happens, as an example, 
we show in Figs.~\ref{Sigc0zrpot} 
and~\ref{Sigc0zrbden} the potential strengths and 
baryon density (together with the baryon density of nucleons), respectively  
for $\Sigma_c^0$ in $^{91}_{\Sigma_c^0}$Zr,  
giving incorrect-"1s$_{1/2}(?)$" (notation used in the figures) 
and correct-2s$_{1/2}$ states.
"Pa.+CC." stands for potential for
the effective Pauli blocking plus the channel coupling effects.
By the terminology, incorrect-"1s$_{1/2}(?)$" state, we mean by that 
the obtained solution is incorrect, since  
both the single-particle energy for the $\Sigma_c^0$ and 
the core nucleon energy 
levels are incorrect.
In this case, the obtained single-particle energies for the $\Sigma_c^0$ 
are, -7.4 and -8.1 MeV, respectively for the "1s$_{1/2}(?)$" 
and 2s$_{1/2}$ states.
Looking at the potential strengths in Fig.~\ref{Sigc0zrpot}, the 
difference between the two states are only slight, and nearly 
indistinguishable. 
However, looking at the $\Sigma_c^0$ baryon density (scaled by a factor 5) 
in Fig.~\ref{Sigc0zrbden}, it is clear that both solutions have two nodes, 
thus actually they are 2s$_{1/2}$ states.
The $\Sigma_c^0$ density distributions are very different, because the 
core nucleon density distributions (particularly around 3 fm 
from the center of the hypernucleus) are 
appreciably different due to the incorrect core nucleon single-particle
energy levels. 
The main difference in the 
calculation of the $\Sigma_c$ hypernuclei from that of 
the $\Sigma$ hypernuclei is the bare (effective) mass,  
where mass of the $\Sigma_c$ is much larger, and the wave function 
(if obtained) for the $1s_{1/2}$ is expected to be localized considerably 
in the central region of each $\Sigma_c$ hypernucleus when the isospin 
dependent strong interaction and Coulomb forces are neglected. 
This may also be a consequence of an interplay between the 
isospin dependent force for the $\Sigma_c^0$ (the isospin third 
component is $-1$), and baryonic density in the central region. 
To test this we have repeated the calculation for $^{91}_{\Sigma^0_c}$Zr 
single-particle energies by switching off both the Pauli blocking
and the channel coupling effects simultaneously by setting 
$\tilde{\lambda}_{\Sigma_c} = 0$. 
The results are shown inside the brackets in Table~\ref{spe2}. 
First, the energy levels from $1p_{3/2}$ to $1g_{9/2}$ states 
seem to be correct.  
But again the correct $1s_{1/2}$ state solution is not found, 
in the sense that the lowest single-particle energy for $s$-state obtained 
is -23.0 MeV, and the corresponding wave function has 
two nodes ($2s_{1/2}$ state).
Then, we have looked into  
the behavior of $\rho$ meson field, particularly near the center of 
$^{91}_{\Sigma^0_c}$Zr, which may show some hints. 
Indeed, we have found following, different and  distinct behaviors 
of the $\rho$ meson field between the $2s_{1/2}$ state and 
that for the other (non-$s$) states:

\begin{itemize}
\item For all the non-$s$ states calculated, $\rho$ meson field is repulsive
and varies in the range $+2 \sim +5$ MeV up to $\sim 4$ fm
from the center of $^{91}_{\Sigma^0_c}$Zr, 
and gradually dies out towards the surface.
\item For the $2s_{1/2}$ state, it is strong {\it attraction} relative 
to that for the non-$s$ states, 
$-28 \sim -20$ MeV up to $\sim 0.4$ fm from the center of 
$^{91}_{\Sigma^0_c}$Zr, 
and changes to {\it repulsion} around $\sim 1$ fm, to become 
maximum of $\sim +7$ MeV around $4$ fm from the center,  
and similarly for the other states, 
gradually dies out towards the surface.
\end{itemize}

This implies that the $1s_{1/2}$ state would need much stronger  
attractive $\rho$ meson field than that for the $2s_{1/2}$ state 
to exist near the center of $^{91}_{\Sigma^0_c}$Zr.
However, such a case would certainly disturb  
the core neutrons which are bound deeply nearer to the center of  
$^{91}_{\Sigma^0_c}$Zr, and result in giving incorrect 
core neutron energy levels as we have to discard such a 
solution observed. Therefore we discard such a solution.
Thus, we would conclude that the peculiar feature of not finding 
the correct $1s_{1/2}$ state in $^{91}_{\Sigma^0_c}$Zr
(and probably also for $^{49}_{\Sigma^0_c}$Ca) is due to 
the behavior of the isospin-dependent 
$\rho$ meson field at very near the center of $^{91}_{\Sigma^0_c}$Zr.
If one ignores the incorrect energy levels of the core neutrons,  
one would get a solution and regard it as a "$1s_{1/2}$" state, 
which is not correct in reality.
As we will discuss later again in case of $1s_{1/2}$ state in  
$^{209}_{\Sigma^0_c}$Pb, indeed the isospin dependent 
$\rho$-meson field is expected to work in a peculiar manner 
for the $\Sigma^0_c$-$1s_{1/2}$ state
in neutron rich $\Sigma^0_c$ hypernuclei. 

On the other hand, recent experiment~\cite{Sigmaexpt} 
suggests that a strongly repulsive $\Sigma$-nucleus potential with a 
non-zero size of the imaginary part is favorable to reproduce the 
measured spectra for middle and large baryon number hypernuclei. 
If we wish to be consistent with this result, 
we need to introduce even stronger repulsive potential,  
either by the channel coupling effect, or based on the 
quark model analysis to reproduce the $\Sigma$-atom data  
without resorting to the channel coupling effect~\cite{Sigma_atom}, 
for the treatments of both the $\Sigma$ and $\Sigma_c$ 
hypernuclei. 
In this case, the repulsion should be strong enough 
so that it entails no bound state 
for the $\Sigma$ hypernuclei (and probably also for the $\Sigma_c$ 
hypernuclei) in the middle and large baryon number hypernuclei. 
This will need a more elaborate investigation 
in the future, using further accumulated experimental data and analyses.

\begin{figure}[hbt]
\begin{center}
\epsfig{file=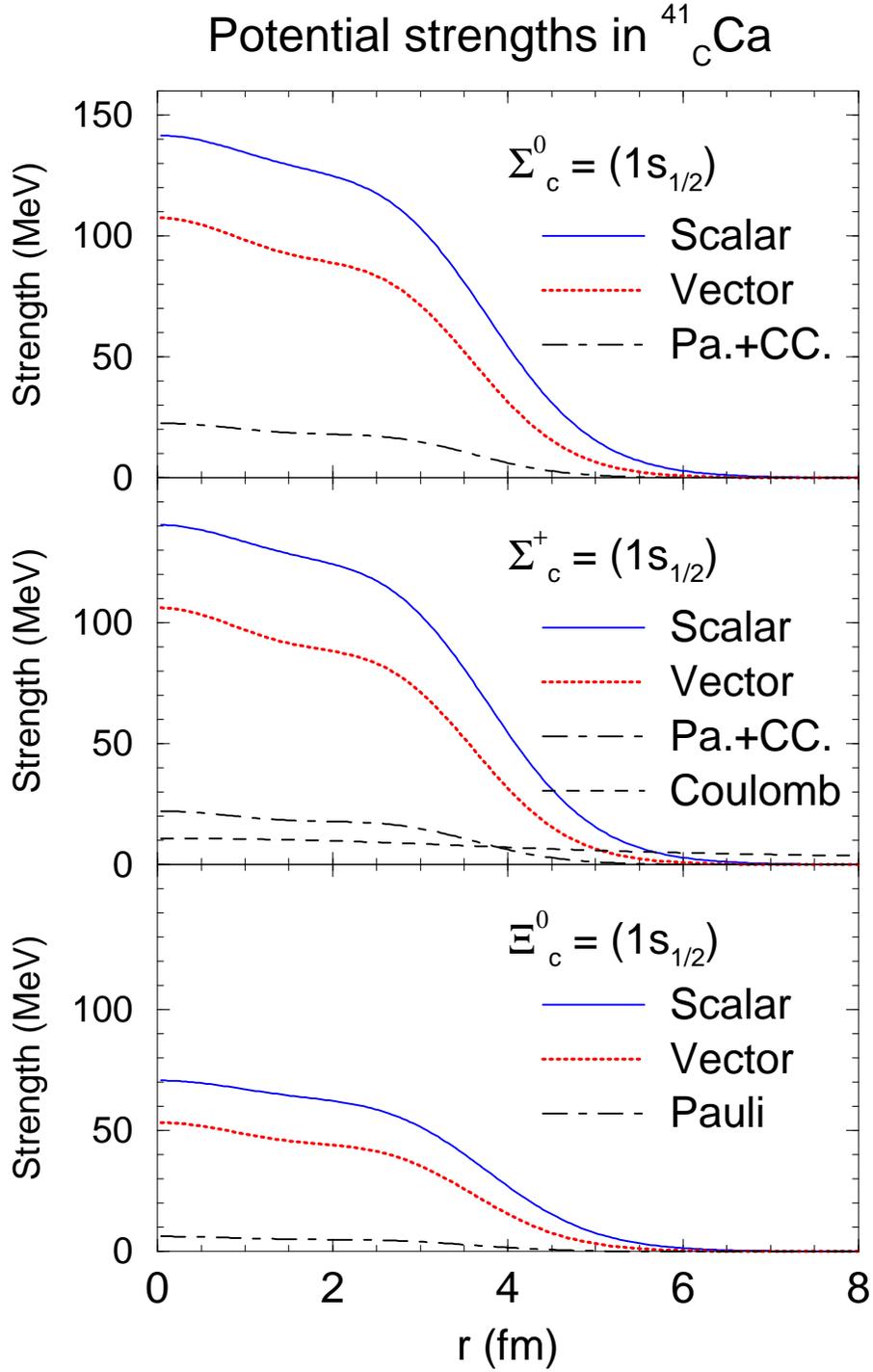,height=12cm,angle=-90}
\caption{
Potential strengths for the $\Sigma_c^{0,+}$ and $\Xi_c^0$ 
in $^{41}_j$Ca ($j = \Sigma_c^{0,+},\Xi_c^0$) for
the $1s_{1/2}$ state. 
See also caption of Fig.~\protect\ref{Sigc0zrpot}. 
\label{CaSXpot}
}
\end{center}
\end{figure}

\begin{figure}[hbt]
\begin{center}
\epsfig{file=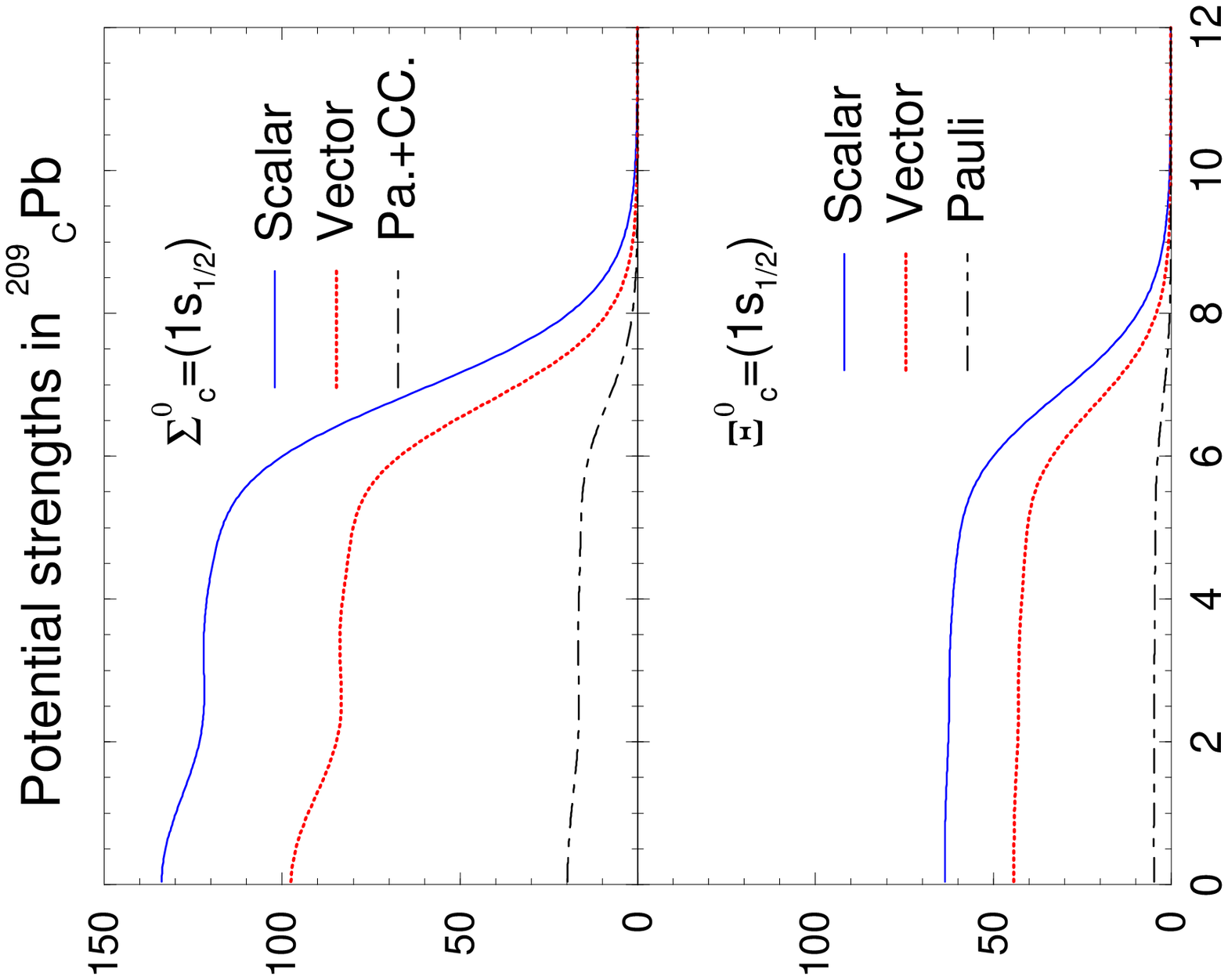,height=12cm,angle=-90}
\caption{
Potential strengths for the $\Sigma_c^0$ and $\Xi_c^0$ in 
$^{209}_j$Pb ($j = \Sigma_c^0,\Xi_c^0$) for the $1s_{1/2}$ state.
See also caption of Fig.~\protect\ref{Sigc0zrpot}. 
\label{PbSXpot}
}
\end{center}
\end{figure}

Next, in Figs.~\ref{CaSXpot} and~\ref{PbSXpot} 
we show the potential strengths 
for the $1s_{1/2}$ state found in  
$^{41}_j$Ca and $^{209}_j$Pb 
($j = \Sigma_c^{0,+},\Xi_c^0$).  
Recall that the effects are different for the $\Lambda_c^+$ and 
$\Sigma_c$ as they are true for the $\Lambda$ and $\Sigma$.
As in the limit of nuclear matter in Section 2.2, the scalar and vector
potentials for the heavy baryons are also quite similar to the 
strange hyperons with the same light quark 
numbers in the corresponding hypernuclei.
This feature also holds between the $\Sigma_c$ and 
$\Sigma$ in the corresponding hypernuclei, except for a  
contributions due to the differences in charges, where the Coulomb force 
affects the baryon density distributions and then  
the vector and scalar potentials are also slightly modified. 
Thus, as far as the total baryon density distributions and
the scalar and vector potentials are concerned, among/between the members 
within each multiplet,
($\Lambda,\Lambda^+_c,\Lambda_b$), ($\Sigma,\Sigma_c$) 
and ($\Xi,\Xi_c$) hypernuclei, they show quite similar features. 
However, in realistic nuclei, the Coulomb force plays a crucial 
role as mentioned before, and the single-particle energies thus obtained 
show very different features within each hypernuclei multiplet.  
Of course, the mass differences within the multiplet is also the 
dominant source for the differences in the obtained 
single-particle energies. 

\begin{figure}[hbt]
\begin{center}
\epsfig{file=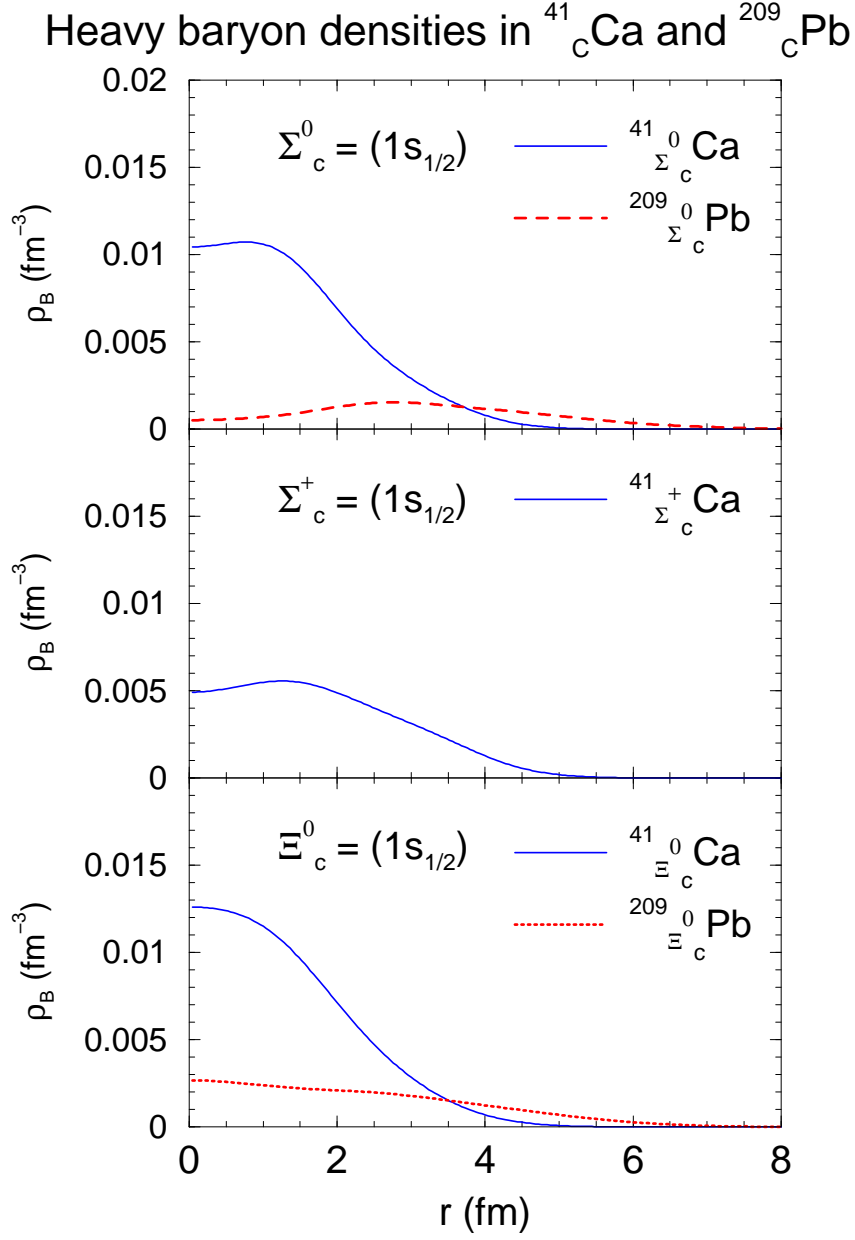,height=12cm,angle=-90}
\caption{$\Sigma_c^{0,+}$ and $\Xi_c^0$
baryon (probability) density distributions for the $1s_{1/2}$ state
in $^{41}_j$Ca and
$^{209}_j$Pb ($j = \Sigma_c^{0,+},\Xi_c^0$).
\label{HBSXdensity}
}
\end{center}
\end{figure}

The probability 
density distributions for the heavy baryons for the $1s_{1/2}$ state, 
in $^{41}_j$Ca and $^{209}_j$Pb ($j = \Sigma_c^{0,+},\Xi_c^0$),  
are shown in Fig.~\ref{HBSXdensity}.
(We mention that the wave function for the $1s_{1/2}$ state 
for $^{41}_{\Sigma^0_c}$Ca is correct, because it has one node as 
shown in Fig.~\ref{HBSXdensity}.)
It is interesting to compare the 
$\Sigma_c^0$ and $\Xi_c^0$ probability density 
distributions.
Due to the Pauli blocking and the channel coupling effects, 
the probability density distributions 
for the $\Sigma_c^0$ are pushed away from the origin 
compared to those for 
the $\Xi_c^0$ in both $^{41}_{\Sigma_c^0}$Ca and $^{209}_{\Sigma_c^0}$Pb.
In particular, the $\Sigma_c^0$ density distributions 
in $^{209}_{\Sigma_c^0}$Pb are really pushed 
away from the central region, and thus nearly losing the character of 
a typical $1s_{1/2}$ state wave function.
This fact shows that the isospin dependent $\rho$-meson mean field
for the $\Sigma^0_c$ $1s_{1/2}$ state in central region of the neutron rich 
nuclei is very important as discussed before, 
and that explains why the correct $1s_{1/2}$ state wave functions are not 
obtained in $^{49}_{\Sigma^0_c}$Ca and $^{91}_{\Sigma^0_c}$Zr, 
where the (isospin asymmetric) baryon density distributions 
in central region of these nuclei are expected to be larger than 
those for the $^{209}_{\Sigma_c^0}$Pb as in the case of normal nuclei. 
(In the case of $^{17}_{\Sigma_c^0}$O, we expect that the size of 
the nucleus is  
much smaller and $\Sigma^0_c$ feels more sensitively the isospin 
dependent $\rho$-meson mean field, due to the higher baryon density 
and the limitation of the mean field approximation than those for 
the $^{209}_{\Sigma_c^0}$Pb.)
On the contrary, it is rather surprising that the $\Xi_c^0$ 
probability density distributions are higher and localized 
in the central region than those for the $\Sigma_c^{0,+}$,  
although one can naively expect that they have an opposite characteristic, 
because of the smaller scalar attractive potential for the $\Xi_c^0$. 
Thus, in the present calculation the effects of  
the Pauli blocking, and particularly the channel coupling, play  
an important role for these different features.
Then, we must be careful in drawing any definite conclusions from 
the $\Sigma_c^0$ and $\Sigma_c^+$ probability density distributions.

%
%
\begin{table}[htbp]
\begin{center}
\caption{Binding energy per baryon, $-E/A$ (in MeV), r.m.s charge radius,
$r_{ch}$, and r.m.s radii of the $\Lambda$ and heavy baryons, $r_j$,
neutron, $r_n$, and proton, $r_p$ (in fm) for $^{17}_j$O, $^{41}_j$Ca, 
$^{49}_j$Ca, $^{91}_j$Zr and $^{209}_j$Pb 
($j = \Lambda, \Lambda^+_c, \Sigma_c, \Xi_c, \Lambda_b$). 
The configurations of the $\Lambda$ and heavy baryon 
$j$, are $1s_{1/2}$ for all hypernuclei.
}
\label{prmst}
\begin{tabular}[t]{cccccc}
\\
\hline \hline
hypernuclei & $-E/A$ & $r_{ch}$ & $r_j$ & $r_n$ & $r_p$ \\
\hline \hline
$^{17}_\Lambda$O        &6.37&2.84&2.49&2.59&2.72\\
$^{17}_{\Lambda^+_c}$O  &6.42&2.85&2.19&2.58&2.73\\
$^{17}_{\Sigma^+_c}$O   &6.10&2.83&2.57&2.60&2.71\\
$^{17}_{\Xi^0_c}$O      &6.01&2.81&2.34&2.61&2.69\\
$^{17}_{\Xi^+_c}$O      &5.84&2.81&2.70&2.61&2.69\\
$^{17}_{\Lambda_b}$O    &6.69&2.87&1.81&2.57&2.75\\
\hline \hline
$^{41}_\Lambda$Ca       &7.58&3.51&2.81&3.31&3.42\\
$^{41}_{\Lambda^+_c}$Ca &7.58&3.51&2.66&3.31&3.42\\
$^{41}_{\Sigma^0_c}$Ca  &7.54&3.51&2.79&3.31&3.42\\
$^{41}_{\Sigma^+_c}$Ca  &7.42&3.50&3.21&3.31&3.41\\
$^{41}_{\Xi^0_c}$Ca     &7.43&3.49&2.76&3.32&3.40\\
$^{41}_{\Xi^+_c}$Ca     &7.32&3.49&3.11&3.32&3.39\\
$^{41}_{\Lambda_b}$Ca   &7.72&3.53&2.22&3.30&3.43\\
\hline \hline
$^{49}_\Lambda$Ca         &7.58&3.54&2.84&3.63&3.45\\
$^{49}_{\Lambda^+_c}$Ca   &7.54&3.54&2.67&3.63&3.45\\
$^{49}_{\Sigma^+_c}$Ca    &7.39&3.54&3.17&3.64&3.44\\
$^{49}_{\Sigma^{++}_c}$Ca &6.29&3.57&3.62&3.71&3.47\\
$^{49}_{\Xi^0_c}$Ca       &7.34&3.53&2.57&3.65&3.43\\
$^{49}_{\Xi^+_c}$Ca       &7.32&3.53&3.19&3.65&3.43\\
$^{49}_{\Lambda_b}$Ca     &7.64&3.56&2.16&3.63&3.46\\
\hline \hline
$^{91}_\Lambda$Zr       &7.95&4.29&3.25&4.29&4.21\\
$^{91}_{\Lambda^+_c}$Zr &7.85&4.29&3.34&4.30&4.21\\
$^{91}_{\Sigma^+_c}$Zr  &7.77&4.29&4.03&4.30&4.21\\
$^{91}_{\Xi^0_c}$Zr     &7.78&4.28&3.01&4.30&4.20\\
$^{91}_{\Lambda_b}$Zr   &7.93&4.30&2.63&4.29&4.22\\
\hline \hline
$^{209}_\Lambda$Pb       &7.35&5.49&3.99&5.67&5.43\\
$^{209}_{\Lambda^+_c}$Pb &7.26&5.49&4.74&5.68&5.43\\
$^{209}_{\Sigma^0_c}$Pb  &6.95&5.51&4.88&5.70&5.45\\
$^{209}_{\Xi^0_c}$Pb     &7.24&5.49&4.44&5.68&5.43\\
$^{209}_{\Lambda_b}$Pb   &7.48&5.49&3.45&5.66&5.43\\
\hline \hline
\end{tabular}
\end{center}
\end{table}
%

Finally, we show in Table~\ref{prmst}
the calculated binding energy per baryon, $-E/A$,
r.m.s charge radius, $r_{ch}$, and r.m.s radii of 
the heavy baryon (also $\Lambda$), neutron
and proton distributions, $r_j$, $r_n$ and $r_p$, respectively.
The results listed in Table~\ref{prmst} are calculated with
the $1s_{1/2}$ heavy baryon configuration for all cases.
At a first glance, it is very clear that the r.m.s radius for the
$\Lambda_b$ is very small compared to those for other baryons
within the same baryon number hypernuclei as one would expect,
due to heavy mass. On the other hand, $r_{ch}$ and/or $r_p$ 
for the $\Lambda_b$ hypernuclei are usually the largest in the 
all hypernuclei calculated. This implies that the protons in the 
core nucleus are relatively more pushed away compared to the other 
hypernuclei, although such feature is not seen for $r_n$.
The radii, $r_{ch}$, $r_n$ and $r_p$, 
may be grouped in similar magnitudes 
for all heavy baryon hypernuclei and $\Lambda$ hypernuclei 
within the same baryon number multiplet 
($^{41}_j$Ca and $^{49}_j$Ca in Table~\ref{prmst} may be grouped 
together), reflecting the fact that
the effect of the embedded baryon on these
quantities are of order,  
$\simeq M_{C,\Lambda}/[(A-1)M_N+M_{C,\Lambda}]$. 
As for the binding energy per baryon, the energy of 
$\Lambda_b$ hypernuclei is usually the largest among the
same baryon number hypernuclei.
One of the largest contributions for this is the single-particle
energy of the $1s_{1/2}$ state, even after dividing by the total
baryon number. 

%
\section{Summary and discussion}

In summary, we have completed systematic studies of the 
$\Lambda_c^+, \Sigma_c, \Xi_c$ and $\Lambda_b$ hypernuclei 
in the QMC model,  
extending the study made for strange hypernuclei~\cite{Tsushima_hyp}.
The spin-orbit potentials for the heavy baryon hypernuclei 
are negligible for the single-particle energies, because of their 
heavy masses. Our results suggest that the formation of 
$\Lambda_c^+, \Xi_c^0$ and $\Lambda_b$ hypernuclei is expected  
to be quite likely, while that of the $\Sigma_c^{++}$ and $\Xi_c^+$ 
is unlikely. For the $\Sigma_c^0$ and $\Sigma_c^+$ hypernuclei, 
although there may be some possibilities to be formed, 
it is difficult to draw a solid conclusion in view of the 
uncertainties in the model and approximations made in the calculations. 

For the $\Sigma_c$ and $\Xi_c$ hypernuclei,  
the Coulomb force is crucial in the possibility to form the hypernuclei, 
in particular for the $\Sigma_c^{+,++}$ hypernuclei. 
The results for the $\Sigma_c$ hypernuclei imply that 
the phenomenologically introduced channel coupling effects  
may be overestimated when using the 
same strength as that for the $\Sigma$ hypernuclei. 
Thus, we need to investigate further how such effects 
can be included consistently with the underlying quark structure. 
In this sense, some of the results for the $\Sigma_c$ hypernuclei 
need to be taken with caution.

Although there can be numerous speculations
on the implications of the present results we would like to
emphasize that our calculations indicate that 
the $\Lambda^+_c,\Xi_c^0$, and $\Lambda_b$ 
hypernuclei would exist in realistic experimental conditions, 
but there may be lesser possibilities for 
the $\Sigma_c^0$ and $\Sigma_c^+$ hypernuclei. 
Furthermore, it is very unlikely that the $\Sigma_c^{++}$ and $\Xi_c^+$ 
hypernuclei will be formed. 
Experiments at facilities like JHF would provide
quantitative input to gain a better understanding of the
interactions of heavy baryons with nuclear
matter. 
Experiments at colliders such as RHIC, LHC and
Fermilab could provide additional data to establish the
formation and decay of such heavy baryon hypernuclei. A combination of
these data inputs and a careful analysis, with the present
calculations being considered as a first step, would give
a valuable information about the physical implications for
the presence of heavy quarks in finite nuclei or dense nuclear matter.
Furthermore, the experiments would indicate certainly the shortcomings 
of the QMC model and in addition would provide a guide to a proper 
and consistent model to treat baryons with heavy quarks 
in finite nuclei or nuclear matter.
Eventually we hope that these results will have important
implications for studies in astrophysics of neutron stars
and nuclear matter at very high densities. 

As for the other aspects of interests for the properties of heavy 
baryons in nuclear medium, additional studies are needed to 
investigate the semi-leptonic weak decay of 
$\Lambda^+_c,\Sigma_c,\Xi_c$ and $\Lambda_b$ hyperons in nuclear medium.
This will provide information on the weak current form factors of 
heavy baryons in nuclear medium, whether or not the 
axial coupling constant of such heavy system 
will be modified appreciably, as is the case implied for 
the quenching $g_A$ of nucleon, possibly due to the internal 
structure change of the heavy baryon in medium~\cite{qmcgA}.
To determine the role of Pauli blocking and density in influencing 
the decay rates
as compared to the free heavy baryons would be highly desirable.
Such studies can provide important information 
on the hadronization of the quark-gluon
plasma and the transport of hadrons in nuclear matter at high density.
Will the high density lead to a slower decay and a higher
probability to survive its passage through the material ?
At present the study of the presence of
heavy baryons in finite nuclei and/or nuclear matter is 
in its infancy.
Careful investigations, both theoretical and experimental, would lead to
a much better understanding of the properties of heavy quarks 
and/or heavy baryons in finite nuclei and nuclear matter.

\vspace{0.5cm}

\noindent{\bf Acknowledgment}\\
The authors would like to thank Prof. A.W. Thomas for the hospitality
at CSSM, Adelaide, where this work was initiated.
KT acknowledges support and warm hospitality at University of
Alberta, where main part of the calculation was completed.
He also would like to thank Profs. F.S. Navarra and M. Nielsen 
for making it possible to use the computer 
facility in Uni. S\~{a}o Paulo to perform new calculations.
KT was supported by the Forschungszentrum-J\"{u}lich,
contract No. 41445282 (COSY-058), and is supported by 
FAPESP contract 2003/06814-8. 
The work of FK is supported by NSERCC.

%

%
%
\newpage
\newpage
\newpage
\newpage
\newpage
\newpage
\end{document}